\documentclass[a4paper,12pt]{article}

\usepackage{graphics}
\usepackage{latexsym}
\newcommand{\ap}{\alpha^{\prime}}

\newcommand{\cC}{\mathcal{C}}
\newcommand{\cD}{\mathcal{D}}

\newcommand{\cI}{\mathcal{I}}

\newcommand{\cO}{\mathcal{O}}

\newcommand{\cQ}{\mathcal{Q}}
\newcommand{\cR}{\mathcal{R}}

\newcommand{\cT}{\mathcal{T}}

\newcommand{\aaru}{\mathbf{R}}
\newcommand{\shii}{\mathbf{C}}

\newcommand{\sectiono}[1]{\section{#1}\setcounter{equation}{0}}

\begin{document}
\begin{titlepage}
\thispagestyle{empty}
\begin{flushright}
YITP-01-91 \\
UT-979 \\
hep-th/0112169 \\
December, 2001
\end{flushright}

\vskip 1.5 cm

\begin{center}
\noindent{\textbf{\LARGE{CFT Description of Identity String Field: \vspace{0.5cm}\\ 
Toward Derivation of the VSFT Action }}}
\vskip 1.5cm
\noindent{\large{Isao Kishimoto\footnote{E-mail: ikishimo@yukawa.kyoto-u.ac.jp} \ 
and \ Kazuki Ohmori\footnote{E-mail: ohmori@hep-th.phys.s.u-tokyo.ac.jp}}} \\
\vspace{0.8cm}
\noindent{\it ${}^1$Yukawa Institute for Theoretical Physics, Kyoto University}\\ \vspace{2mm}
{\it  Kyoto 606-8502, Japan} \\ \vspace{6mm}
\noindent{\normalsize{\textit{${}^2$Department of Physics, Faculty of Science, University of 
Tokyo}} \\ \vspace{2mm}
\normalsize{\textit{Hongo 7-3-1, Bunkyo-ku, Tokyo 113-0033, Japan}}}
\end{center}
\vspace{1cm}
\begin{abstract}
We concretely define the identity string field as a surface state and deal with it consistently 
in terms of conformal field theory language, never using its formal properties nor 
oscillator representation of it. The generalized gluing and resmoothing theorem provides us 
with a powerful computational tool which fits into our framework. Among others, we can prove 
that in some situations the identity state defined this way actually behaves itself like 
an identity element under the $*$-product. We use these CFT techniques 
to give an explicit expression of the classical solution in the ordinary cubic string field theory 
having the property that the conjectured vacuum string field theory action arises when 
the cubic action is expanded around it. 
\end{abstract}
\end{titlepage}
\newpage
\baselineskip=6mm


\sectiono{Introduction}
In the last one year, vacuum string field theory (VSFT), which was first proposed by 
Rastelli, Sen and Zwiebach in~\cite{RSZ1} as a candidate for the theory which describes 
the open string dynamics around the tachyon vacuum, has intensively been studied. 
In the early days, it was recognized~\cite{RSZ2} that the matter part of the equation 
of motion reduced to an equation satisfied by any projection operator if we 
assume that the classical solution representing a D-brane factorized into the ghost part 
and the matter part, because the kinetic operator $\cQ$ of VSFT was supposed to be made 
purely out of ghost fields. Such projectors, which are squared to itself under the $*$-product 
of string field theory, have been constructed algebraically as `sliver states'~\cite{KP,RSZ2} 
in the matter sector. Without any knowledge of the ghost sector, the ratios of tensions of 
classical sliver solutions with different dimensions were calculated and were shown to agree 
numerically with the known results about the ratios of D-brane tensions. After that, the 
sliver states were reconstructed in terms of boundary conformal field theory~\cite{RSZ4,VSFT,
Matsuo1} and were interpreted as rank-one projection `operators' in the sense that 
they operated in the space of half-string functionals~\cite{split}. 
\smallskip

Research in the ghost sector has started with the work of Hata and Kawano~\cite{HK}, 
who showed that it was possible to determine the form of $\cQ$ uniquely by requiring the 
existence of the Siegel gauge solution. This work, written in terms of operator language, 
was refined by one of us~\cite{Kishimoto} and Okuyama~\cite{Okuyama}. Another choice of $\cQ$ 
was proposed by Gaiotto, Rastelli, Sen and Zwiebach in~\cite{GRSZ}, where they considered 
a local ghost insertion at the midpoint of the open string. Surprisingly, it was also 
shown~\cite{GRSZ} that these two candidates for $\cQ$, which appeared to have different 
origins, coincided numerically with each other. Although the construction of vacuum string 
field theory has been completed this way, we are not allowed to compute the absolute values 
of tensions of the brane solutions\footnote{
Hata and Moriyama~\cite{HM} followed the strategy proposed in~\cite{HK}
that one could determine the tension of the brane solution
by calculating the open string coupling constant $g_o$ defined
as the on-shell scattering amplitude of three `tachyons' (so-called HK state)
living on the brane solution, and suggested that the classical solution
found in~\cite{HK} 
corresponded to the configuration of two D25-branes from the energetic
aspects. Later, it was pointed out in~\cite{RSZ6} that this claim was not valid on
the ground that the authors of~\cite{HK,HM} had used the equation of motion
for their `tachyon state' which in fact had not been satisfied in a strong sense.} 
because this theory arises as such a singular limit that 
it needs some regularization for this purpose. Moreover, it seems difficult to reproduce the known 
physical spectrum of D-branes from the brane solutions in vacuum string field theory. 
Even though we could specify the `on-shell tachyon state' with mass $m^2=-1/\ap$~\cite{HK,RSZ6}, 
it would be a hard task to impose the transversality conditions on the vector (tensor) fields, 
\textit{e.g.} $k^{\mu}A_{\mu}(k)=0$, if we start from the kinetic operator $\cQ$ which is completely 
independent of the matter sector and the factorized solutions. 
\medskip

Confronted with these difficulties, we have been led to consider the solutions which contain 
the BRST operator $Q_B$ or something in them. In fact, such attempts were made in the purely 
cubic formulation of string field theory\footnote{Recently Matsuo~\cite{Matsuo3} reconsidered 
this theory and generalized it to a matrix version to let it be able to describe an 
arbitrary number of D-branes.}~\cite{pcsft} about 15 years ago. There, Horowitz \textit{et.al.} 
showed that it was possible to construct solutions of the form\footnote{Takahashi and Tanimoto 
have also considered such solutions in their recent works~\cite{TT}, though their solutions are not 
related to the problem of tachyon condensation.} $Q_L(f)\cI+c_L(g)\cI$ in the original 
cubic open string field theory (OSFT) such that there were no physical open string excitations 
around them. However, their arguments crucially depended on the formal properties of the 
identity string field $\cI$, which made the arguments less persuasive. 
\medskip

In this paper, we will consistently deal with the identity string field in terms of the conformal 
field theory language, never relying on any formal property such as $\cI *A=A*\cI =A$ for every $A$. 
The generalized gluing and resmoothing theorem allows us to carry out concrete calculations 
fully within our framework. We will show that the identity string field $\cI$ defined this way 
actually acts as the identity element of the $*$-algebra in some situations. Thus, one of the aims 
of this paper is to establish how to treat the identity state using conformal field theory 
in a definite manner.  

Furthermore, this computational scheme will be used to propose a possible form of the tachyon 
vacuum solution. More precisely, we consider a specific configuration of the string field 
including the identity state and the BRST operator, and show that this configuration 
indeed solves the equation 
of motion derived from the ordinary cubic string field theory action defined on a D25-brane. 
Expanded around this solution, the action takes the same form as the conjectured vacuum string 
field theory action! Hence we guess that our solution corresponds to the tachyon vacuum in some 
singular frame. We further discuss that the deformation of this solution may lead to a regularized 
version of the VSFT action postulated in~\cite{GRSZ}, though not so conclusive. 
\medskip

This paper is organized as follows. In section~\ref{sec:3}, we explain our definitions of 
string field theory vertices and of identity string field in the CFT language, which will 
be used throughout the rest of this paper. In section~\ref{sec:4}, we describe our main 
computational tool, namely the generalized gluing and resmoothing theorem, and derive a 
general formula for gluing functions. Section~\ref{sec:5} is devoted to technical calculations 
used later. The readers who are interested only in the relation between VSFT and OSFT can skip 
sections~\ref{sec:4} and \ref{sec:5}, but should not forget that our calculations have been done 
based consistently on the CFT prescription. In section~\ref{sec:VSFT} we propose a solution 
which connects OSFT to VSFT. In section~\ref{sec:disc} we discuss some open problems and present 
possible future directions. 

\sectiono{Our Framework}\label{sec:3}
\subsection{Conformal field theory approach to string field theory}
In this paper we will adopt the conformal field theory description of the string field 
theory vertices~\cite{LPP}. In this formulation, the BPZ inner product and the 3-string vertex 
are defined as 
\begin{eqnarray}
\langle A,B\rangle &=&\left\langle I\circ A(0)\ B(0)\right\rangle_{\mathrm{UHP}}, \label{eq:P} \\
\langle A,B*C\rangle &=& \left\langle f_1^{(3)}\circ A(0)\ f_2^{(3)}\circ B(0)\ f_3^{(3)}\circ 
C(0)\right\rangle_{\mathrm{UHP}}, \label{eq:Q}
\end{eqnarray}
respectively, where $f\circ \cO (0)$ denotes the conformal transform of the vertex operator 
$\cO (0)$ by the conformal map $f$, $\langle\ldots\rangle_{\mathrm{UHP}}$ is the correlation 
function among the vertex operators evaluated on an upper half-plane (UHP), and the 
conformal maps are defined by 
\begin{eqnarray}
I(z)&=&-\frac{1}{z}, \qquad (\mathrm{inversion}) \label{eq:R} \\
f_1^{(3)}(z)&=&h^{-1}\left(e^{-\frac{2\pi i}{3}}h(z)^{\frac{2}{3}}\right), \label{eq:S}\\
f_2^{(3)}(z)&=&h^{-1}\left(h(z)^{\frac{2}{3}}\right), \label{eq:T}\\
f_3^{(3)}(z)&=&h^{-1}\left(e^{\frac{2\pi i}{3}}h(z)^{\frac{2}{3}}\right), \label{eq:U}
\end{eqnarray}
where we have for convenience defined an $SL(2,\shii)$ map $h(z)$ through 
\begin{equation}
h(z)=\frac{1+iz}{1-iz}, \quad h^{-1}(z)=-i\frac{z-1}{z+1}. \label{eq:V}
\end{equation}
$h$ takes an upper half-plane $(\mathrm{Im}\ z\ge 0)$ to the inside of a unit disk $(|h(z)|\le 1)$ 
in a one-to-one way. The state-operator isomorphism has been used to map any state $|\cO\rangle$ 
to the corresponding vertex operator $\cO(0)$ (with a slight abuse of notation) via the relation 
$|\cO\rangle =\cO(0)|0\rangle$, where $|0\rangle$ denotes the $SL(2,\aaru)$-invariant vacuum. 
More detailed explanations are found in~\cite{LPP,Rev}. 

\subsection{Defining the `identity' string field}
\begin{figure}[htbp]
	\begin{center}
	\scalebox{0.8}[0.8]{\includegraphics{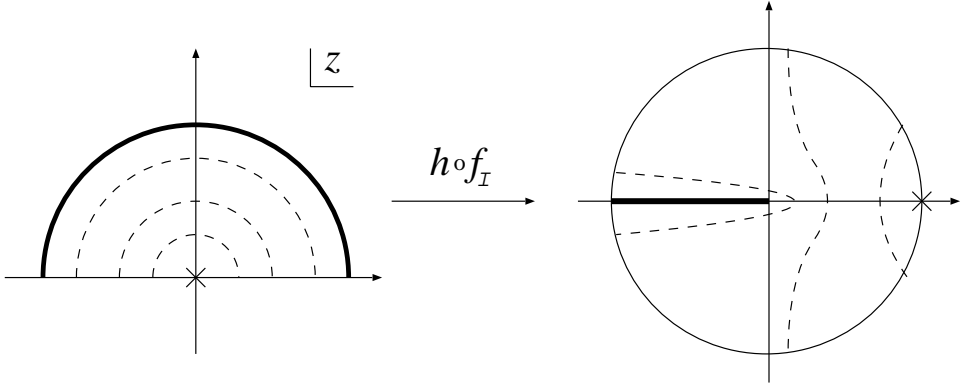}}
	\end{center}
	\caption{Realization of the left-right overlap by means of a conformal map $h\circ 
	f_{\cI}$.}
	\label{fig:id}
\end{figure}

Instead of taking the state $|\cI\rangle$ 
to be an identity element of the $*$-algebra of string field theory, 
we will throughout define it by the relation 
\begin{equation}
\langle\cI, \cO\rangle =\langle f_{\cI}\circ\cO(0)\rangle_{\mathrm{UHP}}, \label{eq:W}
\end{equation}
where 
\begin{equation}
f_{\cI}(z)=h^{-1}\left( h(z)^2\right), \label{eq:X}
\end{equation}
in terms of the conformal field theory language. The reason why the state $\langle\cI |$ 
defined this way may be considered as the `identity' element of open string field theory 
is that the conformal map $f_{\cI}(z)$ geometrically realizes the overlap of left- and 
right-halves of the open string (Fig.\ref{fig:id}), originally referred to as the integration 
operation $\int\Phi$ for an open string field in~\cite{Witten}.
Then it perfectly makes sense to ask whether 
or not $\cI$ really behaves itself like the identity element under the $*$-multiplication. 
In fact, we will show in section~\ref{sec:5} that $\cI *A=A*\cI =A$ certainly holds true for 
a large class of states $A$. 
\medskip

In general, a wedge state $\langle n|$ of an angle $\frac{2\pi}{n}$ was defined in~\cite{RZ} as 
\begin{equation}
\langle n,\cO\rangle =\left\langle f^{(n)}\circ \cO(0)\right\rangle_{\mathrm{UHP}}, \label{eq:Y}
\end{equation}
with
\begin{equation}
f^{(n)}(z)=h^{-1}\left(h(z)^{\frac{2}{n}}\right). \label{eq:Z}
\end{equation}
Hence, our definition of $\langle\cI |$ coincides with that of the wedge state $\langle n=1|$ 
of an angle $2\pi$. 
\medskip

For convenience, here we collect some results which will be obtained in section~\ref{sec:5}.
Below, $\langle\phi |$ and $|\psi\rangle$ denote arbitrary Fock space states, \textit{i.e.} 
those which can be created by the action of local vertex operators on the $SL(2,\aaru)$-invariant 
vacuum as $\langle\phi |=\langle 0|I\circ\phi(0)$, $|\psi\rangle=\psi(0)|0\rangle$. 
$*$-multiplication formulae including $|\cI\rangle$ are:
\begin{eqnarray}
\langle\phi, \cI *\psi\rangle &=&\langle\phi,\psi *\cI\rangle =\langle \phi,\psi\rangle, \label{eq:AA}\\
\langle\phi, \cI*\cO\cI\rangle &=&\langle\phi,\cO\cI*\cI\rangle =\langle\phi,\cO\cI\rangle, 
\label{eq:AB}
\end{eqnarray}
where $\cO$ represents a local vertex operator or a contour-integrate of it. From these equations, 
$\cI$ actually looks like an identity element of the $*$-algebra when we restrict ourselves to 
the above settings. The BRST operator $Q_B$, which governs the perturbative behavior of open 
strings living on a D25-brane, satisfies the following relations: 
\begin{eqnarray}
& &\{ Q_B,Q_L\} =0, \label{eq:AC} \\
& &\langle \cI |Q_B|\psi\rangle=\langle\cI |(Q_R+Q_L)|\psi\rangle=0, \label{eq:AD} \\
& &\langle\phi,(Q_RA)*B+(-1)^{|A|}A*(Q_LB)\rangle=0, \label{eq:AE}
\end{eqnarray}
which have been at the heart of the manipulations in purely cubic string field theory~\cite{pcsft}. 
$Q_L,Q_R$ will be defined in section~\ref{sec:5}. Eq.(\ref{eq:AE}) holds for $A,B$ in a class 
larger than that consisting of the Fock space states.  

\sectiono{Generalized Gluing and Resmoothing Theorem}\label{sec:4}
In the CFT formulation of string field theory, it is helpful to make use of so-called 
generalized gluing and resmoothing theorem (GGRT), which has been developed 
in~\cite{LPP,SchSen,David}, as a computational scheme. After reviewing the geometrical aspects of 
the gluing theorem following~\cite{SchSen} in subsection~\ref{subsec:41}, we will derive a general formula 
for conformal mappings from the original surfaces to the sewed surface, paying great attention to 
the problem of precisely defining the range of angles. 

\subsection{Geometrical description of the sewing procedure}\label{subsec:41}
\begin{figure}[htbp]
	\begin{center}
	\scalebox{0.9}[0.9]{\includegraphics{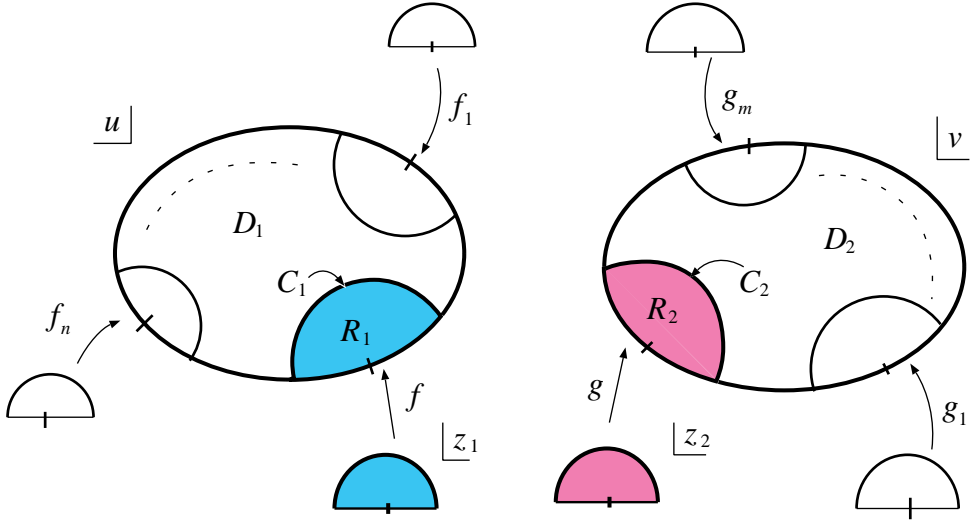}}
	\end{center}
	\caption{Conformal maps $f$'s and $g$'s from upper half-disks to global disks $\cD_1,\cD_2$.}
	\label{fig:maps}
\end{figure}

\begin{figure}[htbp]
	\begin{center}
	\scalebox{0.85}[0.8]{\includegraphics{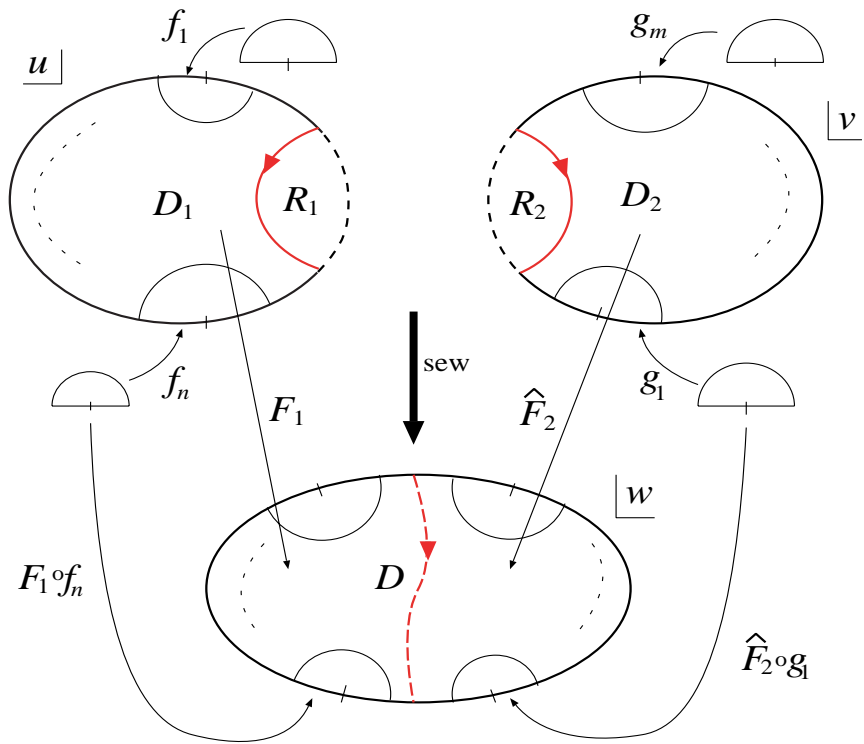}}
	\end{center}
	\caption{A new disk $\cD$ is obtained by sewing the regions $\cD_1-R_1$ and $\cD_2-R_2$ together.}
	\label{fig:sew}
\end{figure}

\begin{figure}[htbp]
	\begin{center}
	\scalebox{0.8}[0.8]{\includegraphics{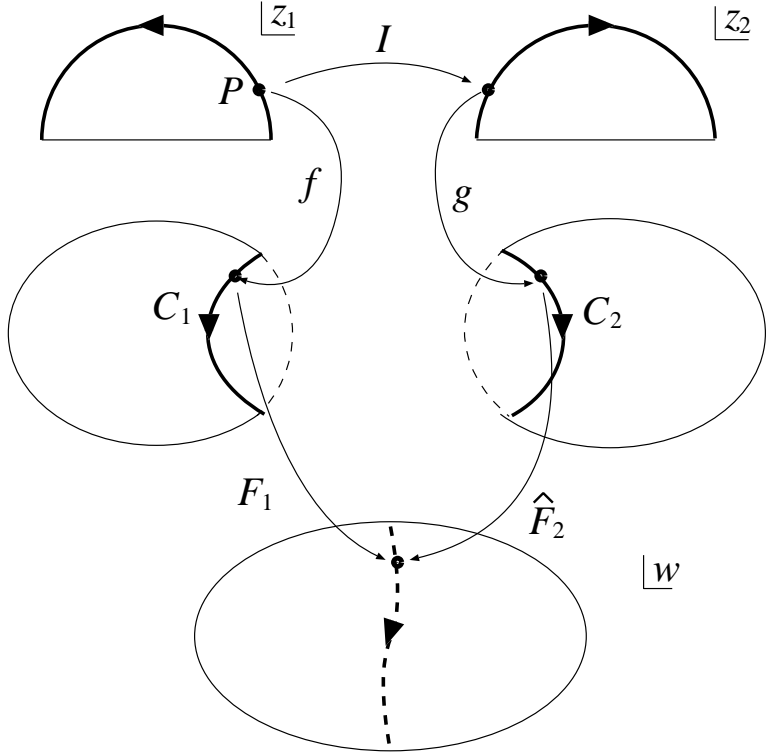}}
	\end{center}
	\caption{The images of a point $P$ under the maps $F_1\circ f$ and $\widehat{F}_2\circ g\circ I$
	must agree in the $w$-plane.}
	\label{fig:coin}
\end{figure}

Let us suppose that we are given $(n+1)$- and $(m+1)$-point correlation functions on 
disks\footnote{Similar arguments can be given for the case of spheres.} $\cD_1,\cD_2$, respectively, 
defined by 
\begin{eqnarray}
I_{\cD_1}^{(n+1)}&=&\langle f_1\circ\Phi_{r_1}(0)\ldots f_n\circ\Phi_{r_n}(0)\ f\circ\Phi_r(0)
\rangle_{\cD_1}, \label{eq:BA} \\
I_{\cD_2}^{(m+1)}&=&\langle g_1\circ\Phi_{s_1}(0)\ldots g_m\circ\Phi_{s_m}(0)\ g\circ\Phi_s(0)
\rangle_{\cD_2}, \label{eq:BB} 
\end{eqnarray}
where $f$'s and $g$'s represent conformal maps each of which embeds a unit upper half-disk defining 
a local coordinate system into the corresponding region in $\cD_1$ or $\cD_2$ (as shown in 
Fig.~\ref{fig:maps}), and $\Phi$'s are local vertex operators defined on their own local coordinate 
disks. 
Now consider sewing two disks $\cD_1$ and $\cD_2$ into another disk $\cD$ in the following way. 
First remove the regions $R_1$ and $R_2$ which are the images of the unit upper half-disks under the 
maps $f$ and $g$, respectively, as indicated in Fig.~\ref{fig:maps}. Let $z_1,z_2$ be the coordinates 
on the unit half-disks corresponding to the regions $f^{-1}(R_1)$, $g^{-1}(R_2)$. 
Then, sew two surfaces $\cD_1-R_1$ and $\cD_2-R_2$ along the curves $C_1$ and $C_2$ $(C_1=\{ f(z_1);
|z_1|=1,\mathrm{Im}z_1\ge 0\}, C_2=\{ g(z_2); |z_2|=1,\mathrm{Im}z_2\ge 0\} )$ through the relation 
\begin{equation}
z_1=I(z_2)=-\frac{1}{z_2}. \label{eq:BC}
\end{equation}
Let us call the sewed surface (disk) $\cD$, and the global coordinate on it $w$. The above sewing 
procedure is shown in Fig.~\ref{fig:sew}.
There we have defined $F_1$ and $\widehat{F}_2$ to be the maps such that $w=F_1(u)$ and 
$w=\widehat{F}_2(v)$, where $u$ and $v$ are global coordinates on $\cD_1$ and $\cD_2$, respectively. 
Looking at the resulting surface $\cD$, one may notice that this surface defines an $(n+m)$-point 
correlation function
\begin{equation}
I_{\cD}^{(n+m)}=\left\langle F_1\circ f_1\circ\Phi_{r_1}(0)\ldots F_1\circ f_n\circ\Phi_{r_n}(0)\ 
\widehat{F}_2\circ g_1\circ\Phi_{s_1}(0)\ldots\widehat{F}_2\circ g_m\circ\Phi_{s_m}(0)
\right\rangle_{\cD}. \label{eq:BD}
\end{equation}
In fact, the generalized gluing and resmoothing theorem states that the correlation function 
$I_{\cD}^{(n+m)}$ on the sewed surface $\cD$ can indeed be obtained by contracting the correlation 
functions $I_{\cD_1}^{(n+1)}, I_{\cD_2}^{(m+1)}$ on the original surfaces $\cD_1,\cD_2$ using 
the `metric' 
\begin{equation}
h^{rs}=\langle\Phi^c_r, \Phi^c_s\rangle=\langle I\circ\Phi^c_r(0)\ \Phi^c_s(0)\rangle \label{eq:BE}
\end{equation}
as 
\begin{equation}
I^{(n+m)}_{\cD}=\sum_{r,s}I^{(n+1)}_{\cD_1}(\Phi_{r_1},\ldots ,\Phi_{r_n},\Phi_r)
I^{(m+1)}_{\cD_2}(\Phi_{s_1},\ldots ,\Phi_{s_m},\Phi_s)h^{rs}. \label{eq:BF}
\end{equation}
In eq.(\ref{eq:BE}), $\{\langle\Phi^c_r|\}$ denotes the set of bra-states which is dual to the 
complete set $\{|\Phi_r\rangle\}$ of ket-states in the sense that 
\begin{equation}
\langle\Phi_r^c,\Phi_s\rangle=\delta_{rs}. \label{eq:BG2}
\end{equation}
More explicitly, we have 
\begin{eqnarray}
& &\sum_r\left\langle f_1\circ\Phi_{r_1}(0)\ldots f_n\circ\Phi_{r_n}(0)\ f\circ\Phi_r(0)
\right\rangle_{\cD_1}\left\langle g_1\circ\Phi_{s_1}(0)\ldots g_m\circ\Phi_{s_m}(0)\ g\circ\Phi_r^c(0)
\right\rangle_{\cD_2} \nonumber \\
&=&\left\langle F_1\circ f_1\circ\Phi_{r_1}(0)\ldots F_1\circ f_n\circ\Phi_{r_n}(0)\ 
\widehat{F}_2\circ g_1\circ\Phi_{s_1}(0)\ldots\widehat{F}_2\circ g_m\circ\Phi_{s_m}(0)
\right\rangle_{\cD}, \label{eq:BG}
\end{eqnarray}
for a theory with vanishing total central charge. Before completing the gluing procedure, 
we have to construct the conformal mappings $F_1$ and $\widehat{F}_2$ explicitly. 
Since $\cD_1-R_1$ and $\cD_2-R_2$ are glued together along the curves $C_1$ and $C_2$ with 
the identification~(\ref{eq:BC}), the image of a point $P(z_1)$ on the arc $\{|z_1|=1,\mathrm{Im}
z_1\ge 0\}$ under the map $F_1\circ f$ must coincide with the image of the corresponding point 
$z_2=-1/z_1$ under the map $\widehat{F}_2\circ g$ in the $w$-plane (Fig.~\ref{fig:coin}). 
This condition is expressed as 
\begin{equation}
F_1\circ f(z)=\widehat{F}_2\circ g\circ I(z) \label{eq:BH}
\end{equation}
around the joint curve $\{|z|=1,\ \mathrm{Im}z\ge 0\}$. In the next subsection, we shall explain how to 
find such functions $F_1,\widehat{F}_2$ for given $f,g$ of the special form. 

\subsection{Constructing $F_1$ and $\widehat{F}_2$ for wedges}\label{subsec:construct}
For the case of string field theory, we often encounter the situation in which the conformal maps 
$f$ and $g$ take the form 
\begin{equation}
f(z)=h^{-1}\left(e^{i\eta_1}h(z)^{\gamma_1}\right),\quad g(z)=h^{-1}\left( e^{i\eta_2}h(z)^{\gamma_2}
\right), \label{eq:BI}
\end{equation}
where $h(z)$ is defined in~(\ref{eq:V}). The function $h\circ f$ maps a unit upper half-disk to a wedge $R_1$ 
of an angle $\pi\gamma_1$ bounded by a unit circle. The global interaction disk $\cD_1$ 
(to be more precise we should write it as $h(\cD_1)$, but we will omit $h$ below) in this case 
is represented by a unit disk. Hence the region $\cD_1-R_1$ complementary to the wedge $R_1$ in $\cD_1$ 
also takes the form of a wedge of an angle $2\pi-\pi\gamma_1$. The same as above holds for the regions 
$\cD_2,R_2$ if we replace $f,\gamma_1$ by $g,\gamma_2$, respectively. According to the strategy 
described in the last subsection, we will sew these two wedges $\cD_1-R_1$ and $\cD_2-R_2$ together 
to make up a new unit disk $\cD$ with no conical singularities. The r\^ole of this sewing procedure 
is played by the conformal maps $F_1$ and $\widehat{F}_2$. Since both of $\cD_1-R_1$ and $\cD_2-R_2$ 
are wedges, we expect that desired maps $F_1,\widehat{F}_2$ can be obtained by combining the rigid 
rotations of the wedges around the origin with the changes of angles of the wedges. Hence 
we put the following ansatz 
\begin{equation}
F_1(z)=h^{-1}\left(e^{i\phi}h(z)^{\alpha}\right),\quad \widehat{F}_2(z)=h^{-1}\left( e^{i\theta}
h(z)^{\beta}\right), \label{eq:BJ}
\end{equation}
and determine the values of $\alpha,\beta,\phi,\theta$ in such a way that the eq.(\ref{eq:BH}) 
should be satisfied. 
\medskip

\begin{figure}[htbp]
	\begin{center}
	\scalebox{0.8}[0.8]{\includegraphics{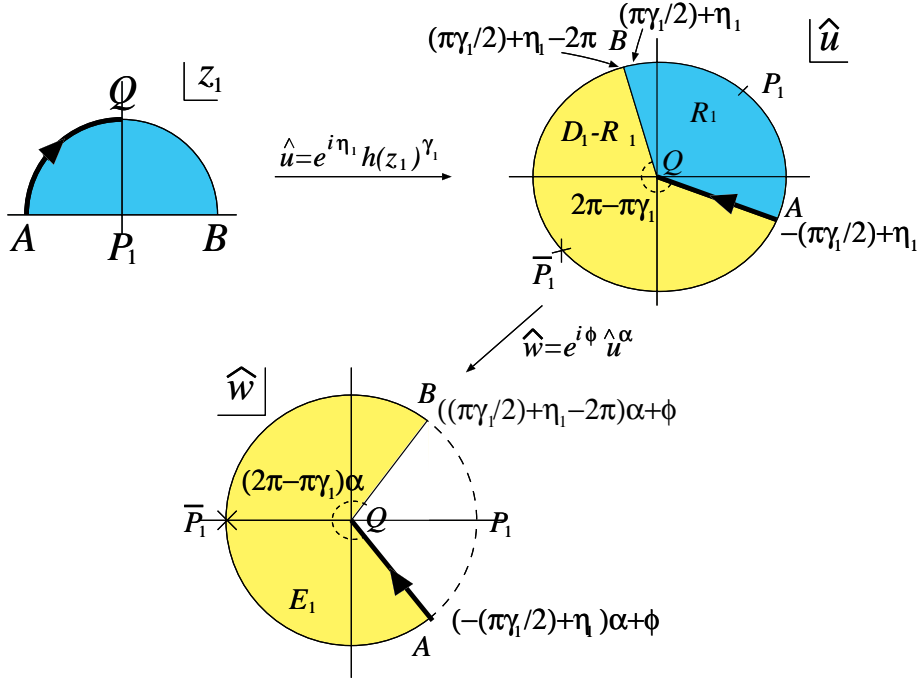}}
	\end{center}
	\caption{Geometrical construction of the map $F_1\circ f(z)$.}
	\label{fig:F1f}
\end{figure}

First we consider the map $F_1\circ f(z_1)=h^{-1}(e^{i\phi}(e^{i\eta_1}h(z_1)^{\gamma_1})^{\alpha}
)$, referring to Fig.~\ref{fig:F1f}.
In the following, we will be watching how the left half $QA$ of the string (indicated by a bold arrow) 
moves. Under the map $\widehat{u}=e^{i\eta_1}h(z_1)^{\gamma_1}$, the local coordinate disk 
represented by a unit upper half-disk in the $z_1$-plane is mapped to the region $R_1$ in the 
$\widehat{u}$-plane, and left half-string $QA$ is mapped to a line segment the angle of which is 
$-\frac{\pi}{2}\gamma_1+\eta_1$ to the positive real axis. When we come to consider taking the 
complement of $R_1$, ambiguity begins to arise in the definition of angles. Since the difference 
$2\pi$ in angle on the $\widehat{u}$-plane is expanded (or compressed) to $2\pi \alpha$ on the 
$\widehat{w}$-plane (defined below), this ambiguity could affect the results seriously. We will 
follow the convention that the angle of the line segment $QA$ is common to the two regions $R_1$ and 
$\cD_1-R_1$, whereas the angle of the line segment $QB$ seen from the complement $\cD_1-R_1$ is 
diminished by $2\pi$ as compared to that of $QB$ seen from $R_1$ (as shown in Fig.~\ref{fig:F1f}). 
Then the range $\cD_1-R_1$ of definition of the map $F_1\circ h^{-1}$ 
is unambiguously fixed to the sector 
bounded by the two lines of angles $\left(-\frac{\pi}{2}\gamma_1+\eta_1\right)$ and $\left(\frac{\pi}{2}
\gamma_1+\eta_1-2\pi\right)$ and by a unit circle $|\widehat{u}|=1$. The angle $\angle AQB$ 
of this sector is given by $2\pi-\pi\gamma_1$. Let us mark the point in the middle of the arc $AB$ 
in $\cD_1-R_1$ as $\overline{P_1}$, which is diametrically opposite to the insertion point $P_1$ 
on $\cD_1$ of the local vertex operator $\Phi_r$ associated with the map $f$ in eq.(\ref{eq:BA}). 
As the next step, we perform the mapping $\widehat{w}=e^{i\phi}\widehat{u}^{\alpha}$. 
Making use of the degree of freedom of $\phi$, we take $\overline{P_1}$ to the point $e^{-i\pi}$ 
on the $\widehat{w}$-plane. This procedure of fixing the phase of the point $\overline{P_1}$ is 
not essential for the gluing process because only the \textit{relative} position between two 
regions $\cD_1-R_1$ and $\cD_2-R_2$ on the $\widehat{w}$-plane becomes important. 
Nonetheless, we have done it because it slightly facilitates the treatment of angles. In fact, 
this additional degree of freedom of rotating the unit disk can always be provided by the 
$SL(2,\aaru)$ transformation inside the correlator~(\ref{eq:BD}) so that it obviously 
makes no difference to the final results. Let us call the image of the region $\cD_1-R_1$ 
under the map $\widehat{u}\mapsto\widehat{w}$ the region $E_1$. The geometrical data 
we have obtained are:
\begin{eqnarray}
\mbox{The angle of the sector }E_1 &:& (2\pi-\pi\gamma_1)\alpha, \label{eq:BK} \\
\mbox{The angle of the line segment }QA &:& \left(-\frac{\pi}{2}\gamma_1+\eta_1\right)\alpha
+\phi\in [ -\pi,0], \label{eq:BL} \\
\mbox{The angle of the line segment }QB &:& \left(\frac{\pi}{2}\gamma_1+\eta_1-2\pi\right)\alpha
+\phi\in [ -2\pi,-\pi]. \label{eq:BM}
\end{eqnarray}
The restrictions on the angles follow from the choice $\widehat{w}(\overline{P_1})=e^{-\pi i}$. 
Moreover, it would be obvious that the angle $\angle AQP_1$ is equal to the angle $\angle BQP_1$. 
This condition is algebraically written as 
\begin{equation}
\left(\frac{\pi}{2}\gamma_1+\eta_1-2\pi\right)\alpha+\phi+2\pi=-\left\{\left(-\frac{\pi}{2}+\eta_1\right)
\alpha+\phi\right\}, \label{eq:BN}
\end{equation}
which could be understood if one remembers the definition of angles on the $\widehat{w}$-plane. 
\smallskip

In the same way as above, we construct the other map $\widehat{F}_2\circ g\circ I(z)=h^{-1}
(e^{i\theta}(e^{i\eta_2}h(I(z))^{\gamma_2})^{\beta})$ following Fig.~\ref{fig:F2g}. 
\begin{figure}[htbp]
	\begin{center}
	\scalebox{0.8}[0.8]{\includegraphics{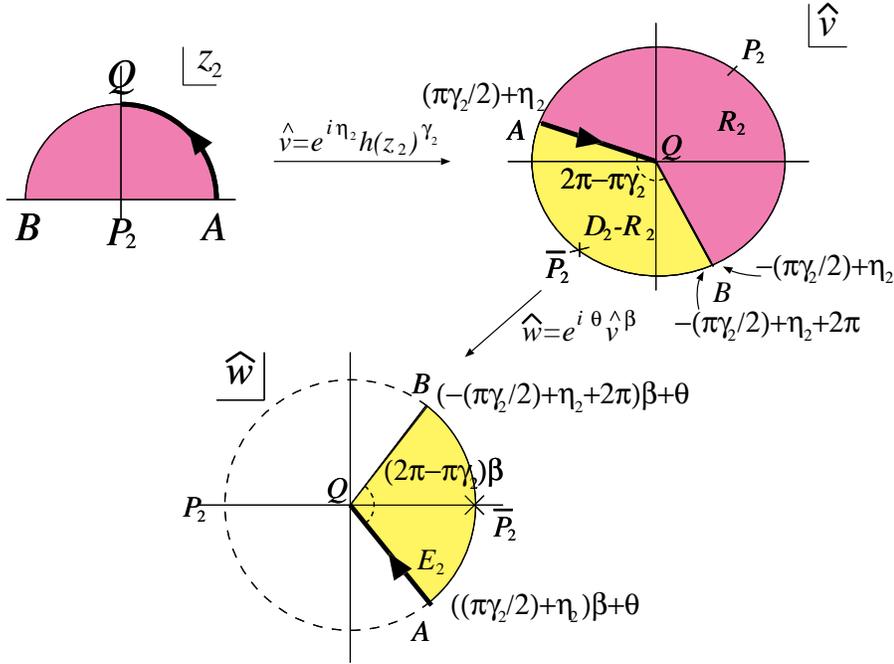}}
	\end{center}
	\caption{Geometrical construction of the map $\widehat{F}_2\circ g\circ I(z)$.}
	\label{fig:F2g}
\end{figure}
Since the gluing will be performed with the identification $z_2=I(z_1)=-1/z_1$, left and right 
of the string should be reversed on the arc $\{ |z_2|=1,\mathrm{Im}z_2\ge 0\}$ as compared to 
the previous case ($z_1$-plane). Under the map $\widehat{v}=e^{i\eta_2}h(z_2)^{\gamma_2}$, 
the unit upper half-disk is mapped to the region $R_2$. In the same sense as in the case of 
the disk $\cD_1$, we keep fixed the angle of the line segment $QA$ to the positive real axis 
during the process of taking the complement of $R_2$ in $\cD_2$. Then the angle of $QB$ seen from 
the region $\cD_2-R_2$ has become $-\frac{\pi}{2}\gamma_2+\eta_2+2\pi$. The point $\overline{P_2}$ 
is defined to be the one in the middle of the arc $AB$ in $\cD_2-R_2$. 
The region $\cD_2-R_2$ is further mapped by $\widehat{w}=e^{i\theta}\widehat{v}^{\beta}$ into 
the region $E_2$ in the $\widehat{w}$-plane, in such a way that $\overline{P_2}$ is now 
taken to the point $e^{0i}$. The set of data drawn from the above consideration is:
\begin{eqnarray}
\mbox{The angle of the sector }E_2 &:& (2\pi-\pi\gamma_2)\beta, \label{eq:BO} \\
\mbox{The angle of the line segment }QA &:& \left(\frac{\pi}{2}\gamma_2+\eta_2\right)\beta
+\theta \in [ -\pi,0], \label{eq:BP} \\
\mbox{The angle of the line segment }QB &:& \left(-\frac{\pi}{2}\gamma_2+\eta_2+2\pi\right)\beta
+\theta \in [ 0,\pi ], \label{eq:BQ} \\
\mbox{Equality of two angles }\angle AQ\overline{P_2},\angle BQ\overline{P_2} &:& 
\nonumber \\
-\left(\frac{\pi}{2}\gamma_2+\eta_2\right)\beta-\theta&=&\left(-\frac{\pi}{2}\gamma_2+\eta_2+2\pi
\right)\beta+\theta. \label{eq:BR}
\end{eqnarray}
Finally, for the eq.(\ref{eq:BH}) to be true, the exponents of $h$ in both sides must be 
identical. Since 
\begin{eqnarray}
F_1\circ f(z)&=&h^{-1}\left( e^{i(\phi+\eta_1\alpha)}h(z)^{\gamma_1\alpha}\right), \label{eq:BS}\\
\widehat{F}_2\circ g\circ I(z)&=&h^{-1}\left( e^{i(\theta+\eta_2\beta)}h(I(z))^{\gamma_2\beta}\right), 
\nonumber 
\end{eqnarray}
this condition gives
\begin{equation}
\gamma_1\alpha =\gamma_2\beta. \label{eq:BT}
\end{equation}
\smallskip

We have now finished gathering the pieces needed to determine the values of $\alpha,\beta,\phi,
\theta$, so we begin to solve them. In order for two regions $E_1$ and $E_2$ to be precisely sewn 
together and to form a unit disk, their angles given in~(\ref{eq:BK}) and (\ref{eq:BO}) must 
add up to $2\pi$, 
\begin{equation}
(2\pi-\pi\gamma_1)\alpha+(2\pi-\pi\gamma_2)\beta=2\pi. \label{eq:BU}
\end{equation}
Solving (\ref{eq:BU}) and (\ref{eq:BT}) simultaneously, we find 
\begin{equation}
\alpha=\frac{\gamma_2}{\gamma_1+\gamma_2-\gamma_1\gamma_2},\quad \beta=
\frac{\gamma_1}{\gamma_1+\gamma_2-\gamma_1\gamma_2}. \label{eq:BV} 
\end{equation}
Then eqs.(\ref{eq:BN}) and (\ref{eq:BR}) give 
\begin{equation}
\phi=-\pi+\frac{(\pi-\eta_1)\gamma_2}{\gamma_1+\gamma_2-\gamma_1\gamma_2}, \quad 
\theta=-\frac{(\pi+\eta_2)\gamma_1}{\gamma_1+\gamma_2-\gamma_1\gamma_2}. \label{eq:BW}
\end{equation}
Since the above construction guarantees that the line segments $QA,QB$ in $E_1$ should 
overlap on the $\widehat{w}$-plane with the corresponding ones $QA,QB$ in $E_2$, 
their angles must agree for consistency: (\ref{eq:BL})=(\ref{eq:BP}), (\ref{eq:BQ})=
(\ref{eq:BM})$+2\pi$. One can verify that these two equations indeed hold with the 
choices~(\ref{eq:BV}) and (\ref{eq:BW}). Finally we must also examine whether $F_1\circ f(z)
=\widehat{F}_2\circ g\circ I(z)$ is satisfied. From eqs.(\ref{eq:BS}), (\ref{eq:BV}) and 
(\ref{eq:BW}), we have 
\begin{eqnarray}
F_1\circ f(z)&=&h^{-1}\left[ e^{i\pi\frac{\gamma_1(\gamma_2-1)}{\gamma_1+\gamma_2-\gamma_1\gamma_2}}
h(z)^{\frac{\gamma_1\gamma_2}{\gamma_1+\gamma_2-\gamma_1\gamma_2}}\right], \label{eq:BX} \\
\widehat{F}_2\circ g\circ I(z)&=&h^{-1}\left[ e^{-i\pi\frac{\gamma_1}{\gamma_1+\gamma_2
-\gamma_1\gamma_2}}h(I(z))^{\frac{\gamma_1\gamma_2}{\gamma_1+\gamma_2-\gamma_1\gamma_2}}\right]. 
\label{eq:BY}
\end{eqnarray}
In general, the expression~(\ref{eq:BY}) is not well-defined because the discontinuous transformation 
$I(z)$ gives rise to a factor of $-1$ through the relation $h(I(z))=-h(z)$. Throughout this paper 
we precisely define it as 
\begin{equation}
h(I(z))=-h(z)\equiv e^{\pi i}h(z). \label{eq:BZ}
\end{equation}
By this definition, it can easily be seen that $\widehat{F}_2\circ g\circ I(z)$ exactly 
agrees with eq.(\ref{eq:BX}), as desired. 
\smallskip

Substituting (\ref{eq:BV}) and (\ref{eq:BW}) into (\ref{eq:BJ}), 
we have reached the following general formula 
\begin{eqnarray}
F_1(z)&=&h^{-1}\left(\exp\left[ i\left((2n+1)\pi+\frac{(\pi-\eta_1)\gamma_2}{\gamma_1
+\gamma_2-\gamma_1\gamma_2}\right)\right] h(z)^{\frac{\gamma_2}{\gamma_1
+\gamma_2-\gamma_1\gamma_2}}\right), \label{eq:form}\\
\widehat{F}_2(z)&=&h^{-1}\left(\exp\left[ i\left((2n+2)\pi-\frac{(\pi+\eta_2)\gamma_1}{\gamma_1
+\gamma_2-\gamma_1\gamma_2}\right)\right] h(z)^{\frac{\gamma_1}{\gamma_1
+\gamma_2-\gamma_1\gamma_2}}\right), \nonumber
\end{eqnarray}
which solves the problem of finding functions satisfying 
\[ F_1\circ f(z)=\widehat{F}_2\circ g\circ I(z) \]
for $f,g$ of the special form~(\ref{eq:BI}). In writing~(\ref{eq:form}), we have added $2(n+1)\pi$ 
commonly to $\phi$ and $\theta$ to make the formula more flexible. 

\sectiono{On Some Technical Details}\label{sec:5}

\subsection{Does $|\cI\rangle$ act as the identity element?}
To begin with, we want to show that we can prove 
\begin{equation}
\langle\phi,\cI *\psi\rangle=\langle\phi, \psi *\cI\rangle =\langle\phi, \psi\rangle \label{eq:CA}
\end{equation}
from the definition~(\ref{eq:W}) of $\cI$ and the algebraic statement~(\ref{eq:BG}) of the 
gluing theorem. As mentioned in section~\ref{sec:3}, 
we take $\langle\phi |$ and $|\psi\rangle$ to be Fock space states. 
Let us consider the most left hand side of eq.(\ref{eq:CA}). Inserting the complete set of basis 
$\sum |\Phi_r\rangle\langle\Phi^c_r|$, we get 
\begin{eqnarray}
\langle\phi,\cI *\psi\rangle&=&\sum_r\langle\phi,\Phi_r*\psi\rangle\langle\Phi^c_r,\cI\rangle 
\nonumber \\ &=&\sum_r\left\langle f^{(3)}_1\circ\phi(0)\ f^{(3)}_2\circ\Phi_r(0)\ 
f^{(3)}_3\circ\psi(0)\right\rangle\langle f_{\cI}\circ\Phi_r^c(0)\rangle, \label{eq:CB} 
\end{eqnarray}
where use was made of definition~(\ref{eq:Q}) of the 3-string vertex and that~(\ref{eq:W}) of 
$\cI$.\footnote{Both the symmetric property of the BPZ inner product, $\langle A,B\rangle=
\langle I\circ A(0)\ B(0)\rangle=\langle B,A\rangle$, and the cyclic symmetry of the 
3-string vertex immediately follow from the $SL(2,\aaru)$-invariance of the correlation 
functions.} From here on we omit the subscript `UHP' of the correlators for simplicity, 
as long as no misunderstanding could occur. 
Summing over $r$ with the help of the gluing theorem~(\ref{eq:BG}) gives 
\begin{equation}
\langle\phi,\cI *\psi\rangle =\left\langle F_1\circ f^{(3)}_3\circ\psi(0)\ F_1\circ f^{(3)}_1
\circ\phi(0)\right\rangle, \label{eq:CC}
\end{equation}
where $F_1$ must satisfy 
\begin{equation}
F_1\circ f^{(3)}_2(z)=\widehat{F}_2\circ f_{\cI}\circ I(z) \label{eq:CD}
\end{equation}
for $f^{(3)}_2(z)$ given in~(\ref{eq:T}) and $f_{\cI}(z)$ in~(\ref{eq:X}). We can find such 
functions 
\begin{eqnarray}
F_1(z)&=&h^{-1}\left(e^{2(n+1)\pi i+\frac{1}{2}\pi i}h(z)^{\frac{3}{2}}\right), \label{eq:CE} \\
\widehat{F}_2(z)&=&h^{-1}\left( e^{2(n+1)\pi i-\frac{1}{2}\pi i}h(z)^{\frac{1}{2}}\right), \nonumber
\end{eqnarray}
by applying the general formula~(\ref{eq:form}) to this case ($\eta_1=\eta_2=0,\gamma_1=\frac{2}{3},
\gamma_2=2$). Then, we can compute 
\begin{eqnarray}
F_1\circ f^{(3)}_3(z)&=&h^{-1}\left(e^{(2n+2)\pi i+\frac{1}{2}\pi i}\left(e^{-\frac{4\pi i}{3}}
h(z)^{\frac{2}{3}}\right)^{\frac{3}{2}}\right) \nonumber \\
&=&h^{-1}\left( e^{\left(2n+\frac{1}{2}\right)\pi i}h(z)\right)\equiv h^{-1}\circ R_{\pi/2}\circ 
h(z)=\cR_{\pi/2}(z), \label{eq:CF}
\end{eqnarray}
where we have defined $R_{\theta}$ to be the rotation of the complex plane by an angle $\theta$, 
namely $R_{\theta}(z)=e^{i\theta}z$, and $\cR_{\theta}(z)=h^{-1}\circ R_{\theta}\circ h(z)$ which 
belongs to $SL(2,\aaru)$. We have used $\displaystyle f_3^{(3)}(z)=h^{-1}\left( e^{-\frac{4\pi i}{3}}
h(z)^{\frac{2}{3}}\right)$ instead of~(\ref{eq:U}) because the map $F_1\circ h^{-1}$ has been defined 
on the sector ${\cal D}_1-R_1=\{ z;|z|\le 1, -\frac{5}{3}\pi\le\mathrm{arg}z\le -\frac{\pi}{3}\}$, as discussed 
in detail in the last section. If we notice the following relation 
\begin{equation}
h\circ I(z)=h\left(-\frac{1}{z}\right)=\frac{1-i/z}{1+i/z}=-\frac{1+iz}{1-iz}=-h(z)=e^{\pi i}
h(z), \label{eq:CG}
\end{equation}
we have 
\begin{equation}
F_1\circ f^{(3)}_1(z)=h^{-1}\left(e^{2(n+1)\pi i-\frac{1}{2}\pi i}h(z)\right)=h^{-1}\left(
e^{2n\pi i+\frac{1}{2}\pi i}h\circ I(z)\right)=\cR_{\pi/2}\circ I(z). \label{eq:CH}
\end{equation}
Substituting (\ref{eq:CF}) and (\ref{eq:CH}) into (\ref{eq:CC}), we finally obtain\footnote{Note that 
in order for $\langle\phi, \psi\rangle$ to have a non-vanishing value $\phi$ and $\psi$ must have 
the opposite Grassmannality from each other.}
\begin{equation}
\langle\phi,\cI *\psi\rangle=\langle\cR_{\pi/2}\circ\psi(0)\ \cR_{\pi/2}\circ I\circ\phi(0)\rangle
=\langle I\circ\phi(0)\ \psi(0)\rangle=\langle\phi,\psi\rangle, \label{eq:CI}
\end{equation}
which follows from the invariance of the correlator under the $SL(2,\aaru)$-map $\cR_{\pi/2}$. 
In exactly the same way, we can also prove $\langle\phi,\psi *\cI\rangle=\langle\phi,\psi\rangle$. 
\smallskip

Furthermore, we can show 
\begin{equation}
\langle\phi,\cI *\cO\cI\rangle=\langle\phi,\cO\cI *\cI\rangle=\langle\phi, \cO\cI\rangle, \label{eq:CJ}
\end{equation}
where $\cO$ denotes a local vertex operator acting on $|\cI\rangle$, or a contour-integrate of 
such an operator. This relation will play an important r\^ole in section~{\ref{sec:VSFT}}. 
Making use of the gluing theorem twice, we get
\begin{eqnarray}
\langle\phi,\cI *\cO\cI\rangle&=&\sum_{r,s}\langle\phi,\Phi_r*\Phi_s\rangle\langle\Phi_r^c,\cI
\rangle\langle\Phi_s^c,\cO\cI\rangle \nonumber \\
&=&\sum_{r,s}\left\langle f^{(3)}_1\circ\phi(0)\ f^{(3)}_2\circ\Phi_r(0)\ f^{(3)}_3\circ\Phi_s(0)
\right\rangle\langle f_{\cI}\circ\Phi_r^c(0)\rangle\langle f_{\cI}\circ I\circ\cO\ f_{\cI}\circ
\Phi_s^c(0)\rangle \nonumber \\
&=&\sum_s\left\langle F_1\circ f^{(3)}_3\circ\Phi_s(0)\ F_1\circ f^{(3)}_1\circ\phi(0)\right\rangle
\langle f_{\cI}\circ I\circ\cO\ f_{\cI}\circ\Phi_s^c(0)\rangle \nonumber \\
&=&\left\langle G_1\circ F_1\circ f^{(3)}_1\circ\phi(0)\ \widehat{G}_2\circ f_{\cI}\circ I\circ
\cO\right\rangle, \label{eq:CK}
\end{eqnarray}
where 
\begin{eqnarray}
F_1\circ f^{(3)}_2(z)&=&\widehat{F}_2\circ f_{\cI}\circ I(z), \label{eq:CL}\\
G_1\circ \left(F_1\circ f^{(3)}_3\right)(z)&=&\widehat{G}_2\circ f_{\cI}\circ I(z) \label{eq:CM}
\end{eqnarray}
must be satisfied. Since eq.(\ref{eq:CL}) is the same as eq.(\ref{eq:CD}), the answer should 
be given by~(\ref{eq:CE}). The general formula~(\ref{eq:form}) for $f(z)=F_1\circ f^{(3)}_3(z)
=h^{-1}\left(e^{2n\pi i+\frac{1}{2}\pi i}h(z)\right)$ (eq.(\ref{eq:CF})) and $g=f_{\cI}$ gives us 
\begin{eqnarray}
G_1(z)&=&h^{-1}\left(e^{2\pi i(1+n^{\prime}-2n)}h(z)^2\right), \label{eq:CN} \\
\widehat{G}_2(z)&=&h^{-1}\left(e^{\pi i(2n^{\prime}+1)}h(z)\right), \nonumber
\end{eqnarray}
with $n^{\prime}$ some integer. From the relations 
\begin{eqnarray*}
G_1\circ F_1\circ f^{(3)}_1(z)&=&h^{-1}\left(e^{2n^{\prime}\pi i+\pi i}h(z)^2\right)=
\cR_{\pi}\circ f_{\cI}(z), \\
\widehat{G}_2\circ f_{\cI}\circ I(z)&=&h^{-1}\left(e^{2(n^{\prime}+1)\pi i+\pi i}h(z)^2\right)=
\cR_{\pi}\circ f_{\cI}(z), 
\end{eqnarray*}
it follows that 
\begin{equation}
\langle\phi,\cI *\cO\cI\rangle=\langle f_{\cI}\circ\phi(0)\ f_{\cI}\circ\cO\rangle. \label{eq:CO}
\end{equation}
On the other hand, the evaluation of $\langle\phi,\cO\cI\rangle$ leads to 
\begin{equation}
\langle\phi, \cO\cI\rangle=\langle\cI,(I\circ\cO)\phi\rangle=\langle f_{\cI}\circ I\circ\cO\ 
f_{\cI}\circ\phi(0)\rangle=\langle f_{\cI}\circ\phi(0)\ f_{\cI}\circ\cO\rangle \label{eq:CP}
\end{equation}
because 
\begin{equation}
f_{\cI}\circ I(z)=h^{-1}\left(h(I(z))^2\right)=h^{-1}\left((-h(z))^2\right)=h^{-1}(h(z)^2)
=f_{\cI}(z). \label{eq:fif}
\end{equation}
Comparing (\ref{eq:CO}) with (\ref{eq:CP}), we conclude that 
\begin{equation}
\langle\phi, \cI *\cO\cI\rangle=\langle\phi,\cO\cI\rangle, \label{eq:CQ}
\end{equation}
as stated earlier. The remaining side in eq.(\ref{eq:CJ}) can be proven in a similar manner. 
\medskip

Finally we want to evaluate 
\begin{equation}
\langle\phi,\cO_A\cI *\cO_B\cI\rangle. \label{eq:CT}
\end{equation}
Actually we do not need to calculate any more. One can find 
\begin{equation}
\langle\phi,\cO_A\cI *\cO_B\cI\rangle=\langle a_1\circ\phi(0)\ a_2\circ\cO_A\ a_3\circ\cO_B
\rangle,  \label{eq:CU}
\end{equation}
with 
\begin{eqnarray*}
a_1(z)&=&G_1\circ F_1\circ f^{(3)}_1(z)=h^{-1}\left( e^{2n^{\prime}\pi i+\pi i}h(z)^2
\right)=\cR_{\pi}\circ f_{\cI}(z), \\
a_2(z)&=&G_1\circ\widehat{F}_2\circ f_{\cI}\circ I(z)=h^{-1}\left( e^{2\pi i(n^{\prime}+1)
+\pi i}h(z)^2\right)=\cR_{\pi}\circ f_{\cI}(z), \\
a_3(z)&=&\widehat{G}_2\circ f_{\cI}\circ I(z)=h^{-1}\left( e^{2(n^{\prime}+1)\pi i+\pi i}h(z)^2\right)
=\cR_{\pi}\circ f_{\cI}(z),
\end{eqnarray*}
where $F_1,\widehat{F}_2,G_1,\widehat{G}_2$ have already been 
given in eqs.(\ref{eq:CE}) and (\ref{eq:CN}). Now consider the case 
\begin{equation}
\cO_A=\cO_B=\cQ^{\epsilon}=\frac{1}{2i}\left(e^{-i\epsilon}c(ie^{i\epsilon})-e^{i\epsilon}
c(-ie^{-i\epsilon})\right), \label{eq:CV}
\end{equation}
which will appear in section~\ref{sec:VSFT}. In this case eq.(\ref{eq:CU}) reduces to 
\begin{eqnarray}
\langle\phi,\cQ^{\epsilon}\cI *\cQ^{\epsilon}\cI\rangle&=&-\frac{1}{4}\left\langle a_1\circ\phi(0)
\left(\frac{e^{-i\epsilon}}{a_2^{\prime}(ie^{i\epsilon})}c(a_2(ie^{i\epsilon}))-
\frac{e^{i\epsilon}}{a_2^{\prime}(-ie^{-i\epsilon})}c(a_2(-ie^{-i\epsilon}))\right)\right. \nonumber \\
& &{}\left.\times\left(\frac{e^{-i\epsilon}}{a_3^{\prime}(ie^{i\epsilon})}c(a_3(ie^{i\epsilon}))-
\frac{e^{i\epsilon}}{a_3^{\prime}(-ie^{-i\epsilon})}c(a_3(-ie^{-i\epsilon}))\right)\right\rangle. 
\label{eq:CW}
\end{eqnarray}
The fact $a_2(z)=a_3(z)$ means that the two \textit{local fermionic} operators are inserted at the same 
point, which makes the above correlator vanish. One may think that the conformal factors could 
provide infinities, but it does not happen as long as $\epsilon$ is kept finite. Hence we have found that 
\begin{equation}
\langle\phi,\cQ^{\epsilon}\cI *\cQ^{\epsilon}\cI\rangle=0 \qquad \mbox{for arbitrary nonzero }
\epsilon, \label{eq:CX}
\end{equation}
irrespective of the details of the state $|\phi\rangle$. 
\medskip

Here we give a brief comment on the use of the operator formulation of 
string field theory~\cite{operator,GJ}. 
The expression for the identity state there is given by~\cite{GJ}
\begin{equation}
|\cI\rangle=\frac{1}{4i}b^+\left(\frac{\pi}{2}\right)b^-\left(\frac{\pi}{2}\right)\exp\left[\sum_{n=1}^{\infty}(-1)^n\left(-\frac{1}{2n}\alpha_{-n}\cdot  \alpha_{-n} + c_{-n} b_{-n}\right)\right]c_0c_1|0\rangle. \label{eq:CR}
\end{equation}
The $*$-product of string field theory is expressed as 
\[ |A*B\rangle_3={}_1\langle A|{}_2\langle B| |V_3\rangle_{123}, \]
where the 3-string vertex $|V_3\rangle$ is defined in terms of Neumann coefficients $U^{rs}_{nm},X^{rs}_{nm}$ as 
\[ |V_3\rangle_{123}=\exp\left[-\sum_{r,s=1}^3\left(\frac{1}{2}\sum_{n,m\geq 0}a^{(r)\dagger}_n \cdot U^{rs}_{nm}a^{(s)\dagger}_m +\sum_{n\geq 1,m\geq 0}c^{(r)}_{-n} X^{rs}_{nm}b^{(s)}_{-m}\right)\right]|+\rangle_{123} .\]
The BPZ conjugation is implemented by the reflector state 
\[ {}_{12}\langle R|={}_{12}\langle {\tilde +}|\exp\left[-\sum_{n=0}^{\infty}(-1)^na^{(1)}_n\cdot a^{(2)}_n-\sum_{n=1}^{\infty}(-1)^n\left(c^{(1)}_{n}b^{(2)}_{n}+c^{(2)}_{n}b^{(1)}_{n}\right)\right]\left(c^{(1)}_0+c^{(2)}_0\right),\]
\[ |R\rangle_{12}=\left(b^{(1)}_0-b^{(2)}_0\right)\exp\left[-\sum_{n=0}^{\infty}(-1)^na^{(1)\dagger}_n\cdot a^{(2)\dagger}_n+\sum_{n=1}^{\infty}(-1)^n\left(c^{(1)}_{-n}b^{(2)}_{-n}+c^{(2)}_{-n}b^{(1)}_{-n}\right)\right]|+\rangle_{12}. \]
Then, the condition that $\cI *A=A$ holds for any state $A$ is equivalent to the statement 
\begin{equation}
{}_1\langle\cI| |V_3\rangle_{123}=|R\rangle_{23}. \label{eq:CS}
\end{equation}
Although one can show that this indeed holds true in the matter sector up to an overall 
determinant factor, explicit calculations of ${}_1\langle\cI_{\mathrm{g}}| |V_3^{\mathrm{g}}\rangle_{123}$
in the ghost sector give us an expression which looks very different from the reflector~\cite{Kishimoto}.
To make matters worse, na\"{\i}ve computations\footnote{
We used the relations among infinite matrices formally.
} have led us to a strangely-looking result: 
$\cI_{\mathrm{g}}*\cI_{\mathrm{g}}=0$. Of course, since this expression is multiplied by a determinant 
of infinite matrices which might be divergent, it may be possible that the infinite determinant factor 
compensates for the apparent vanishing of $\cI_{\mathrm{g}}*\cI_{\mathrm{g}}$. Even if this would be 
the case, however, there seems to be no natural way of regularizing it. For these depressing results, 
we have given up dealing with the identity state in the operator formalism. 

\subsection{Concerning the BRST operator}\label{subsec:BRST}
We will later give arguments which depend on the special properties possessed by the BRST charge 
$Q_B$ appearing in the original action~(\ref{eq:A}). In this subsection we prepare for them.
\smallskip

\begin{figure}[htbp]
	\begin{center}
	\scalebox{0.7}[0.7]{\includegraphics{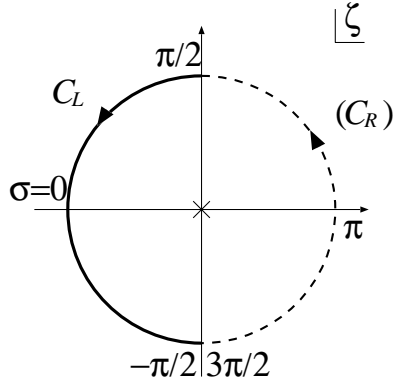}}
	\end{center}
	\caption{The integration contour $C_L$ for $Q_L$.}
	\label{fig:cl}
\end{figure}

We first fix the possible ambiguity of the BRST current $j_B(z)$ by requiring that it be a primary 
field of conformal weight 1. The result is 
\begin{equation}
j_B=cT^{\mathrm{m}}+:bc\partial c:+\frac{3}{2}\partial^2c, \label{eq:CY}
\end{equation}
where $T^{\mathrm{m}}$ denotes the matter part of the energy-momentum tensor. 
Since the OPE of $j_B$ with itself has the single pole proportional to $c^{\mathrm{m}}-26$ with 
$c^{\mathrm{m}}$ denoting the matter central charge, it immediately follows that in the case of 
critical bosonic string theory 
\begin{equation}
\{Q_B,j_B(z)\}=0 \qquad \mbox{for any }z. \label{eq:CZ}
\end{equation}
Here the BRST charge $Q_B$ has been defined by 
\begin{equation}
Q_B=\oint\limits_C\frac{d\zeta}{2\pi i}j_B(\zeta)=-\int\limits_0^{\pi}\frac{d\sigma}{2\pi}
(j_B(\sigma)+\widetilde{\jmath}_B(\sigma)), \label{eq:FA}
\end{equation}
where the integration contour $C$ encircles $z$ counterclockwise, and 
we have used the convention that $j_B(\zeta)$ is a holomorphic field defined over the 
whole complex plane via the doubling trick, whereas $j_B(\sigma),\widetilde{\jmath}_B(\sigma)$ 
are holomorphic and antiholomorphic fields, respectively, defined only on the upper half-plane. 
As a result, we in particular obtain 
\begin{equation}
\{ Q_B,Q_L\}=0, \label{eq:FB}
\end{equation}
where 
\begin{equation}
Q_L\equiv -\int\limits_0^{\pi/2}\frac{d\sigma}{2\pi}(j_B(\sigma)+\widetilde{\jmath}_B(\sigma))
=\int\limits_{C_L}\frac{d\zeta}{2\pi i}j_B(\zeta), \label{eq:FC}
\end{equation}
and the contour $C_L$ is indicated in Fig.~\ref{fig:cl}\footnote{We are following the convention that 
two world-sheet coordinates $(\sigma,\tau)$ and $(z,\bar{z})$ are related by $z=-e^{-i\sigma+\tau},
\bar{z}=-e^{i\sigma+\tau}$.} for $Q_L$ acting on $|\phi\rangle=\phi(0)|0\rangle$. 
We further define $Q_R$ by $Q_B-Q_L$. 
\smallskip

Since $Q_B$ has been defined to be the integral of a primary field of conformal weight 1, 
it commutes with the conformal transformation. This fact can also be seen from $[Q_B,
L_n^{\mathrm{tot}}]=0$ for any $n$. It then follows that the BRST charge annihilates arbitrary 
wedge states $\langle n|$ defined in~(\ref{eq:Y}), (\ref{eq:Z}), because 
\begin{eqnarray}
\langle n|Q_B|\phi\rangle&=&\left\langle f^{(n)}\circ(Q_B\phi)(0)\right\rangle=\left\langle
Q_B\left(f^{(n)}\circ\phi(0)\right)\right\rangle \nonumber \\
&=&\left\langle\oint\limits_C\frac{d\zeta}{2\pi i}j_B(\zeta)\left(f^{(n)}\circ\phi(0)\right)
\right\rangle=0 \label{eq:FD}
\end{eqnarray}
by deforming the contour $C$ until it shrinks to a point at infinity where there is no other 
operator. In particular, we have $\langle\cI |Q_B=0$. 
\medskip

\begin{figure}[htbp]
	\begin{center}
	\scalebox{0.7}[0.7]{\includegraphics{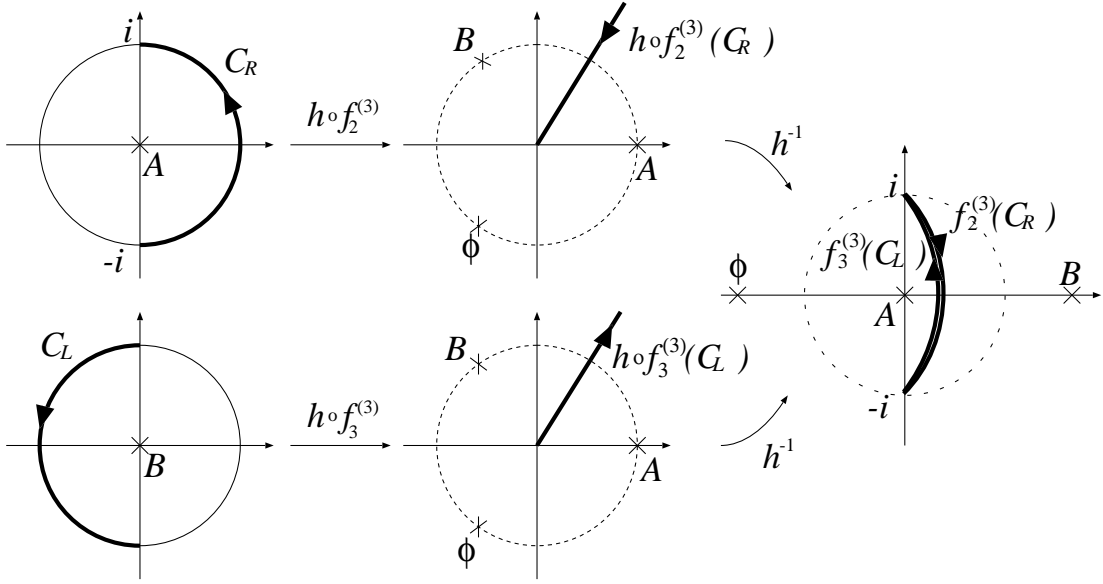}}
	\end{center}
	\caption{The mapping of contours $C_R,C_L$ by $f^{(3)}_2,f^{(3)}_3$.}
	\label{fig:cont}
\end{figure}

An important r\^ole is played in later calculations by a sort of `partial integration formula'
\begin{equation}
\left\langle\phi,(Q_R A)*B\right\rangle=-(-1)^{|A|}\left\langle\phi,A*(Q_L B)
\right\rangle, \label{eq:FE}
\end{equation}
where $|A|$ denotes the Grassmannality of $A$. When both $A$ and $B$ are Fock space states, 
it can be shown simply by considering the integration contours. The contours $C_R$ and $C_L$ 
are mapped under $f^{(3)}_2$ and $f^{(3)}_3$, respectively, as in Fig.~\ref{fig:cont}. 
Then 
\begin{eqnarray}
& &\langle\phi,(Q_R A)*B+(-1)^{|A|}A*(Q_L B)\rangle \nonumber \\
&=&\left\langle f^{(3)}_1\circ\phi(0)\int_{f^{(3)}_2(C_R)}\frac{dz}{2\pi i}j_B(z)\ 
f^{(3)}_2\circ A(0)\ f^{(3)}_3\circ B(0)\right\rangle \nonumber \\ & & {}+
\left\langle f^{(3)}_1\circ\phi(0)\int_{f^{(3)}_3(C_L)}\frac{dz}{2\pi i}j_B(z)\ 
f^{(3)}_2\circ A(0)\ f^{(3)}_3\circ B(0)\right\rangle \nonumber \\
&=&\left\langle f^{(3)}_1\circ\phi(0)\int_{f^{(3)}_2(C_R)+f^{(3)}_3(C_L)}\frac{dz}{2\pi i}
j_B(z)\ f^{(3)}_2\circ A(0)\ f^{(3)}_3\circ B(0)\right\rangle. \label{eq:FF}
\end{eqnarray}
Since the contour $C=f^{(3)}_2(C_R)+f^{(3)}_3(C_L)$ is closed and does not encircle any operator 
inside it, the above expression~(\ref{eq:FF}) vanishes due to the holomorphicity of $j_B(z)$. 
Hence we have shown eq.(\ref{eq:FE}). 
\smallskip

Even when $A$ and $B$ are of the form 
\begin{equation}
|A\rangle=\cO_A|m\rangle,\quad |B\rangle=\cO_B|m\rangle, \label{eq:piwedge}
\end{equation}
where $\cO_A,\cO_B$ represent local operators or contour-integrates of them, and $|m\rangle$ 
is an arbitrary wedge state, we can also give a similar argument to the above one. 
Using the gluing theorem twice, we have obtained 
\begin{eqnarray}
& &\left\langle\phi, (Q_R\cO_A|m\rangle)*\cO_B|m\rangle+(-1)^{|\cO_A|}\cO_A|m\rangle
*(Q_L\cO_B|m\rangle)\right\rangle \nonumber \\
&=&\Biggl\langle G_1\circ F_1\circ f^{(3)}_1\circ\phi(0)\ \int\limits_{\cC}\frac{dz}{2\pi i}
j_B(z)\ G_1\circ\widehat{F}_2\circ f^{(m)}\circ I\circ\cO_A \nonumber \\ 
& &\qquad\qquad\qquad \times \widehat{G}_2\circ f^{(m)}\circ I\circ\cO_B\Biggr\rangle ,
\label{eq:FG}
\end{eqnarray}
where 
\begin{equation}
\cC=G_1\circ\widehat{F}_2\circ f^{(m)}\circ I(C_R)+\widehat{G}_2\circ f^{(m)}\circ I(C_L)
\label{eq:SA}
\end{equation}
and $F_1,\widehat{F}_2,G_1,\widehat{G}_2$ satisfy 
\begin{equation}
F_1\circ f^{(3)}_2(z)=\widehat{F}_2\circ f^{(m)}\circ I(z), \quad G_1\circ \left(F_1\circ 
f^{(3)}_3\right)(z)=\widehat{G}_2\circ f^{(m)}\circ I(z), \label{eq:SB} 
\end{equation}
so these are explicitly given by 
\begin{eqnarray}
F_1(z)&=&h^{-1}\left( e^{i(2n+1)\pi+i\pi\frac{3}{m+1}}h(z)^{\frac{3}{m+1}}\right), \label{eq:SC} \\
\widehat{F}_2(z)&=&h^{-1}\left( e^{2(n+1)\pi i-i\pi\frac{m}{m+1}}h(z)^{\frac{m}{m+1}}
\right), \label{eq:SD} \\
G_1(z)&=&h^{-1}\left( e^{i(2n^{\prime}+1)\pi -2n\pi i\frac{m+1}{2m-1}+\pi i\frac{1}{2m-1}}
h(z)^{\frac{m+1}{2m-1}}\right), \label{eq:SE} \\
\widehat{G}_2(z)&=&h^{-1}\left( e^{2(n^{\prime}+1)\pi i-i\pi\frac{m}{2m-1}}h(z)^{\frac{m}{2m-1}}
\right), \label{eq:SF}
\end{eqnarray}
with $n,n^{\prime}$ integers. We note that 
\begin{eqnarray}
F_1\circ f^{(3)}_3(z)&=&h^{-1}\left( e^{i(2n+1)\pi+i\pi\frac{3}{m+1}}\left( e^{-\frac{4\pi i}{3}}
h(z)^{\frac{2}{3}}\right)^{\frac{3}{m+1}}\right) \nonumber \\
&=&h^{-1}\left( e^{2n\pi i+i\pi\frac{m}{m+1}}h(z)^{\frac{2}{m+1}}\right). \label{eq:SG}
\end{eqnarray}
According to the arguments given in subsection~\ref{subsec:construct}, these conformal maps are 
defined over the regions 
\begin{eqnarray}
F_1\circ h^{-1} &:& \left\{ z; -\frac{5}{3}\pi\le\mathrm{arg}z\le -\frac{\pi}{3}\right\}, \label{eq:SH} \\
\widehat{F}_2\circ h^{-1} &:& \left\{ z; \frac{\pi}{m}\le\mathrm{arg}z\le\frac{2m-1}{m}\pi\right\}, 
\label{eq:SI} \\
G_1\circ h^{-1} &:& \left\{ z; (2n-1)\pi\le \mathrm{arg}z\le 2n\pi+\frac{m-1}{m+1}\pi\right\}, 
\label{eq:SJ} \\
\widehat{G}_2\circ h^{-1} &:& \left\{ z; \frac{\pi}{m}\le\mathrm{arg}z\le\frac{2m-1}{m}\pi
\right\}, \label{eq:SK}
\end{eqnarray}
respectively, where we have not restricted them to the inside $|z|\le 1$ of the unit disk 
because we are considering the full complex plane due to the doubling trick. 
\smallskip

\begin{figure}[htbp]
	\begin{center}
	\scalebox{0.7}[0.7]{\includegraphics{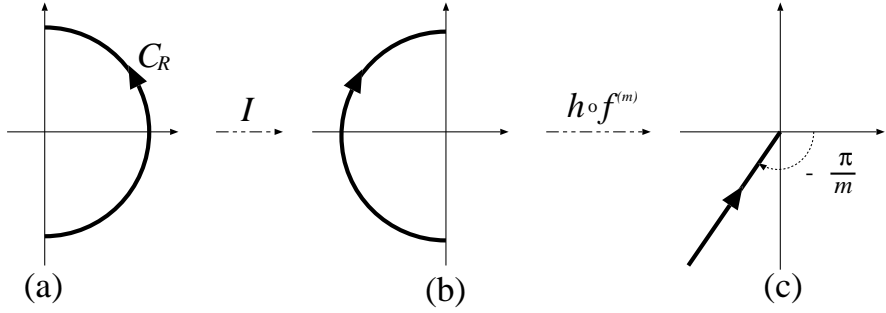}}
	\end{center}
	\caption{The mappings of $C_R$.}
	\label{fig:A1}
\end{figure}

\begin{figure}[htbp]
	\begin{center}
	\scalebox{0.7}[0.7]{\includegraphics{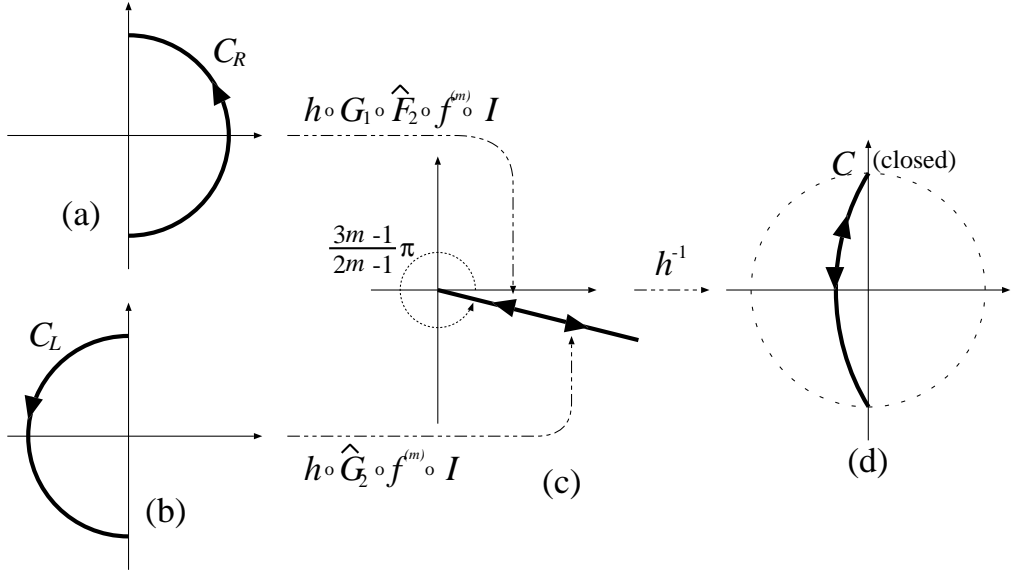}}
	\end{center}
	\caption{The contour $\cC=G_1\circ\widehat{F}_2\circ f^{(m)}\circ I(C_R)
	+\widehat{G}_2\circ f^{(m)}\circ I(C_L)$.}
	\label{fig:A2}
\end{figure}

Now let us see where the integration contours $C_R,C_L$ are mapped under $G_1\circ\widehat{F}_2
\circ f^{(m)}\circ I$ and $\widehat{G}_2\circ f^{(m)}\circ I$, respectively. The contour $C_R$ is first 
mapped by the inversion $I$ to the left semi-circle (Fig.~\ref{fig:A1}(b)), which is subsequently 
mapped under $h\circ f^{(m)}$ into the semi-infinite straight line whose angle is $-\pi/m$ 
to the positive real axis (Fig.~\ref{fig:A1}(c)). 
Since this line lies outside the 
range~(\ref{eq:SI}) of definition of the map $\widehat{F}_2\circ h^{-1}$, we redefine its angle 
as $-\pi/m+2\pi=\pi(2m-1)/m$, which is contained in the domain~(\ref{eq:SI}). This line is mapped 
under $h\circ \widehat{F}_2\circ h^{-1}$~(\ref{eq:SD}) to another semi-infinite straight line with 
an angle 
\begin{equation}
\frac{2m-1}{m}\pi\times\frac{m}{m+1}+2(n+1)\pi-\pi\frac{m}{m+1}=2n\pi+2\pi+\frac{m-1}{m+1}\pi. \label{eq:SL}
\end{equation}
This angle, however, is outside the range~(\ref{eq:SJ}) of definition of $G_1\circ h^{-1}$, so that 
it should be redefined to be $2n\pi+\frac{m-1}{m+1}\pi$. This line is finally mapped by $h\circ G_1\circ
h^{-1}$~(\ref{eq:SE}) to a semi-infinite straight line of an angle 
\begin{eqnarray}
\left( 2n\pi+\frac{m-1}{m+1}\pi\right)\times\frac{m+1}{2m-1}&+&(2n^{\prime}+1)\pi-2n\pi
\frac{m+1}{2m-1}+\frac{\pi}{2m-1} \nonumber \\
&=&2n^{\prime}\pi+\frac{3m-1}{2m-1}\pi. \label{eq:SM}
\end{eqnarray}
Note that an arbitrary integer $n$ has dropped out, and that the remaining one $n^{\prime}$ 
is clearly irrelevant. To summarize, the contour $C_R$ is mapped by $h\circ G_1\circ\widehat{F}_2
\circ f^{(m)}\circ I$ to the semi-infinite straight line of an angle $\frac{3m-1}{2m-1}\pi$ 
to the positive real axis, as indicated in Fig.~\ref{fig:A2}(c). 

In a similar way, we can examine where $C_L$ is mapped under $h\circ\widehat{G}_2\circ
f^{(m)}\circ I$ as 
\begin{equation}
C_L\stackrel{I}{\longrightarrow}-C_R\stackrel{h\circ f^{(m)}}{\longrightarrow}
\left\{ z; \mathrm{arg}z=\frac{\pi}{m}\right\}\stackrel{h\circ\widehat{G}_2\circ h^{-1}}{\longrightarrow}
\left\{ z; \mathrm{arg}z=2n^{\prime}\pi+\frac{3m-1}{2m-1}\pi\right\}, \label{eq:SN}
\end{equation}
which precisely coincides with $\{ h\circ G_1\circ\widehat{F}_2\circ f^{(m)}\circ I(C_R)\}$ 
found earlier in~(\ref{eq:SM}). Moreover, we can easily show that the directions of these two 
transformed contours are opposite to each other (Fig.~\ref{fig:A2}(c)). Mapped by $h^{-1}$, 
the contour $\cC$~(\ref{eq:SA}) becomes a `closed' curve with no room for any operator insertion, 
as shown in Fig.~\ref{fig:A2}(d). Therefore, we have found that the integral of $j_B(z)$ along 
the contour $\cC$ appearing in~(\ref{eq:FG}) vanishes, which means that we can carry out 
the partial integration 
\begin{equation}
\left\langle\phi,(Q_R\cO_A|m\rangle)*\cO_B|m\rangle\right\rangle=-(-1)^{|\cO_A|}\left\langle
\phi,\cO_A|m\rangle*(Q_L\cO_B|m\rangle)\right\rangle \label{eq:SO}
\end{equation}
in the case~(\ref{eq:piwedge}) as well. 
\medskip

We will show that the formula~(\ref{eq:FE}) also holds for the following `mixed' cases, 
\begin{equation}
|A\rangle=\cO_A|m\rangle, \quad |B\rangle=B(0)|0\rangle \label{eq:SP}
\end{equation}
and 
\begin{equation}
|A\rangle=A(0)|0\rangle, \quad |B\rangle=\cO_B|m\rangle. \label{eq:SQ}
\end{equation}
For the former case~(\ref{eq:SP}), we have 
\begin{eqnarray}
& &\left\langle\phi, (Q_R\cO_A|m\rangle)*B+(-1)^{|\cO_A|}\cO_A|m\rangle*(Q_LB)\right\rangle \nonumber \\
&=&\left\langle F_1\circ f_1^{(3)}\circ\phi(0)\int\limits_{\cC^{\ell}}\frac{dz}{2\pi i}j_B(z)\ 
\widehat{F}_2\circ f^{(m)}\circ I\circ \cO_A\ F_1\circ f_3^{(3)}\circ B(0)\right\rangle, \label{eq:SR}
\end{eqnarray}
where $F_1$ and $\widehat{F}_2$ have been given in eqs.(\ref{eq:SC})--(\ref{eq:SD}), and 
\begin{equation}
\cC^{\ell}=\widehat{F}_2\circ f^{(m)}\circ I(C_R) +F_1\circ f^{(3)}_3(C_L). \label{eq:SS}
\end{equation}
In the same spirit as in eq.(\ref{eq:SN}), we follow where the contours $C_R$ and $C_L$ are
mapped to in~(\ref{eq:SS}) as 
\begin{eqnarray}
C_R&\stackrel{I}{\longrightarrow}&-C_L\stackrel{h\circ f^{(m)}}{\longrightarrow}\left\{
z;\mathrm{arg}z=-\frac{\pi}{m}\cong\frac{2m-1}{m}\pi\right\} \nonumber \\
& &\stackrel{h\circ\widehat{F}_2\circ h^{-1}}{\longrightarrow}\left\{z;\mathrm{arg}z
=2(n+1)\pi+\frac{m-1}{m+1}\pi\right\}, \label{eq:ST} \\
C_L&\stackrel{h\circ f^{(3)}_3}{\longrightarrow}&\left\{z;\mathrm{arg}z=\frac{\pi}{3}
\cong -\frac{5}{3}\pi\right\}\stackrel{h\circ F_1\circ h^{-1}}{\longrightarrow}
\left\{z;\mathrm{arg}z=2n\pi+\frac{m-1}{m+1}\pi\right\}, \label{eq:SU}
\end{eqnarray}
where the symbol $\cong$ means that we have added or subtracted $2\pi$ according to the 
ranges of definition of the subsequent maps. Since the $SL(2,\shii)$-map $h^{-1}(z)$ 
to be performed finally does not give rise to any deficit angle, the above two 
contours~(\ref{eq:ST}), (\ref{eq:SU}) precisely overlap with each other in the opposite 
direction on the global complex plane, despite the fact that they differ by $2\pi$ in angle. 
Hence the contour $\cC^{\ell}$ can shrink to zero-size without picking up any poles, 
making the correlator in the right hand side of eq.(\ref{eq:SR}) vanish. So we have proven 
\begin{equation}
\left\langle\phi,(Q_R\cO_A|m\rangle)*B\right\rangle=-(-1)^{|\cO_A|}\left\langle\phi,
\cO_A|m\rangle*(Q_LB)\right\rangle. \label{eq:SV}
\end{equation}
\smallskip

For the latter case~(\ref{eq:SQ}), we go through similar steps: 
\begin{eqnarray}
& &\left\langle\phi,(Q_RA)*\cO_B|m\rangle+(-1)^{|A|}A*(Q_L\cO_B|m\rangle)\right\rangle \nonumber \\
&=&\left\langle F_1^r\circ f_1^{(3)}\circ\phi(0)\int\limits_{\cC^r}\frac{dz}{2\pi i}j_B(z)\ 
F_1^r\circ f_2^{(3)}\circ A(0)\ \widehat{F}_2^r\circ f^{(m)}\circ I\circ\cO_B\right\rangle, \label{eq:SW}
\end{eqnarray}
where 
\begin{eqnarray}
F_1^r(z)&=&h^{-1}\left( e^{2n\pi i+i\pi\frac{m+2}{m+1}}h(z)^{\frac{3}{m+1}}\right);\quad 
\mathrm{arg}h(z)\in\left[ -\pi, \frac{\pi}{3}\right], \label{eq:SX} \\
\widehat{F}_2^r(z)&=&h^{-1}\left( e^{2n\pi i+i\pi\frac{m+2}{m+1}}h(z)^{\frac{m}{m+1}}\right); 
\quad \mathrm{arg}h(z)\in\left[\frac{\pi}{m},\frac{2m-1}{m}\pi\right], \label{eq:SY} \\
\cC^r&=&F_1^r\circ f_2^{(3)}(C_R)+\widehat{F}_2^r\circ f^{(m)}\circ I(C_L). \label{eq:SZ}
\end{eqnarray}
The mappings of the contours $C_L,C_R$ are as follows: 
\begin{eqnarray}
C_R&\stackrel{h\circ f_2^{(3)}}{\longrightarrow}&\left\{z;\mathrm{arg}z=\frac{\pi}{3}\right\}
\stackrel{h\circ F_1^r\circ h^{-1}}{\longrightarrow}\left\{z;\mathrm{arg}z=2n\pi+\frac{m+3}{m+1}\pi
\right\}, \label{eq:RA} \\
C_L&\stackrel{h\circ f^{(m)}\circ I}{\longrightarrow}&\left\{z;\mathrm{arg}z=\frac{\pi}{m}\right\}
\stackrel{h\circ\widehat{F}_2^r\circ h^{-1}}{\longrightarrow}\left\{z;\mathrm{arg}z=2n\pi+
\frac{m+3}{m+1}\pi\right\}. \label{eq:RB}
\end{eqnarray}
Therefore, one again finds $\cC^r\sim \{\mathrm{pt.}\}$, so that 
\begin{equation}
\left\langle\phi,(Q_RA)*\cO_B|m\rangle\right\rangle=-(-1)^{|A|}
\left\langle\phi,A*(Q_L\cO_B|m\rangle)\right\rangle. \label{eq:RC}
\end{equation}
\medskip

We note here some relations which have been established and will be used later. 
\begin{eqnarray}
\langle\phi,\cO\cI*Q_L \cI\rangle&=&-(-1)^{|\cO|}\langle\phi,Q_R\cO\cI *
\cI\rangle, \label{eq:FI} \\
\langle\phi,Q_R\cI * \cO\cI\rangle&=&-\langle\phi,\cI *Q_L\cO\cI\rangle, \label{eq:FJ} \\
\langle\phi,Q_R\cI*\psi\rangle&=&-\langle\phi,\cI *Q_L\psi\rangle, \label{eq:FK} \\
\langle\phi,\psi *Q_L\cI\rangle&=&-(-1)^{|\psi|}\langle\phi,Q_R\psi*\cI\rangle, \label{eq:FL} 
\end{eqnarray}

\subsection{Regularized $\cQ$ as an inner derivation}\label{subsec:cQ}
Now we consider an `inner derivation' of the form
\begin{equation}
\cQ^{\epsilon}\cI*\psi-(-1)^{|\psi|}\psi*\cQ^{\epsilon}\cI, \label{eq:FO}
\end{equation}
where $\cQ^{\epsilon}$ will be defined in~(\ref{eq:O}). 
Taking the BPZ inner product with a Fock space state $|\phi\rangle$ and calculating it with 
the gluing theorem, we obtain 
\begin{eqnarray}
& &\langle\phi, \cQ^{\epsilon}\cI*\psi\rangle-(-1)^{|\psi|}\langle\phi,\psi*
\cQ^{\epsilon}\cI\rangle \nonumber \\
&=&\left\langle F_1\circ f_3^{(3)}\circ\psi(0)\ F_1\circ f_1^{(3)}\circ\phi(0)\ 
\widehat{F}_2\circ f_{\cI}\circ I\circ\cQ^{\epsilon}\right\rangle \label{eq:FP} \\
& &{}-(-1)^{|\psi|}\left\langle F_3\circ f_1^{(3)}\circ\phi(0)\ F_3\circ f_2^{(3)}
\circ\psi(0)\ \widehat{F}_4\circ f_{\cI}\circ I\circ\cQ^{\epsilon}\right\rangle, \nonumber
\end{eqnarray}
where $F$'s satisfy 
\begin{equation}
F_1\circ f_2^{(3)}(z)=\widehat{F}_2\circ f_{\cI}\circ I(z), \quad 
F_3\circ f_3^{(3)}(z)=\widehat{F}_4\circ f_{\cI}\circ I(z). 
\end{equation}
The solution to the above equations is found to be 
\begin{eqnarray*}
F_1(z)&=&h^{-1}\left(e^{2(n+1)\pi i+\frac{1}{2}\pi i}h(z)^{\frac{3}{2}}\right), \quad 
\widehat{F}_2(z)=h^{-1}\left(e^{2(n+1)\pi i-\frac{1}{2}\pi i}h(z)^{\frac{1}{2}}\right), \\
F_3(z)&=&h^{-1}\left(e^{(2n^{\prime}+1)\pi i+\frac{1}{2}\pi i}h(z)^{\frac{3}{2}}\right), \quad 
\widehat{F}_4(z)=h^{-1}\left(e^{2(n^{\prime}+1)\pi i-\frac{1}{2}\pi i}h(z)^{\frac{1}{2}}\right).
\end{eqnarray*}
Then the right hand side of eq.(\ref{eq:FP}) becomes
\begin{eqnarray}
& &\langle\cR_{\pi/2}\circ\psi(0)\ \cR_{\pi/2}\circ I\circ\phi(0)\ \cR_{\pi/2}\circ\cQ^{\epsilon}
\rangle \nonumber \\ & &{}-(-1)^{|\psi|}\langle\cR_{\pi/2}\circ\phi(0)\ \cR_{\pi/2}\circ I\circ
\psi(0)\ \cR_{\pi/2}\circ\cQ^{\epsilon}\rangle \nonumber \\
&=&\langle I\circ\phi(0)\ (\cQ^{\epsilon}-I\circ\cQ^{\epsilon})\psi(0)\rangle =
\langle\phi |(\cQ^{\epsilon}-I\circ\cQ^{\epsilon})|\psi\rangle, \label{eq:FQ}
\end{eqnarray}
where $\cR_{\theta}$ has been defined in~(\ref{eq:CF}). From the definition~(\ref{eq:O}) of 
$\cQ^{\epsilon}$, we can explicitly write down $\cQ^{\epsilon}-I\circ\cQ^{\epsilon}$ as 
\begin{eqnarray}
\cQ^{\epsilon}-I\circ\cQ^{\epsilon}&=&\frac{1}{2i}\biggl(e^{-i\epsilon}c(ie^{i\epsilon})-
e^{i\epsilon}c(-ie^{-i\epsilon})-e^{-i\epsilon}I\circ c(ie^{i\epsilon})+e^{i\epsilon}
I\circ c(-ie^{-i\epsilon})\biggr) \nonumber \\
&=&2\times\frac{1}{4i}\biggl(e^{-i\epsilon}c(ie^{i\epsilon})+
e^{i\epsilon}c(ie^{-i\epsilon})-e^{-i\epsilon}c(-ie^{i\epsilon})-e^{i\epsilon}
c(-ie^{-i\epsilon})\biggr) \nonumber \\
&\equiv&2\cQ^A_{\epsilon}. \label{eq:FR}
\end{eqnarray}
Making use of the operator $\cQ^A_{\epsilon}$ defined above, which na\"{\i}vely seems to approach 
$\frac{\cQ}{(g_o^2\kappa_0)^{1/3}}=\frac{1}{2i}\Bigl( c(i)-c(-i)\Bigr)$ in the 
limit $\epsilon\to 0$ (see eq.(\ref{eq:G})), the expression~(\ref{eq:FO}) can be written as 
\begin{equation}
\frac{1}{2}\langle\phi,\cQ^{\epsilon}\cI*\psi\rangle-\frac{(-1)^{|\psi|}}{2}\langle\phi,\psi
*\cQ^{\epsilon}\cI\rangle=\langle\phi,\cQ^A_{\epsilon}\psi\rangle. \label{eq:FS}
\end{equation}
In fact, this relation has already been stated without proof in eq.(2.27) of~\cite{GRSZ}. 
(Our $\cQ^A_{\epsilon}$ was denoted as $\cQ_{\epsilon}$ there, and their $|S_{\epsilon}\rangle$ 
corresponds to our $\frac{1}{2}\cQ^{\epsilon}|\cI\rangle$.) It can easily be seen that this 
$\cQ^A_{\epsilon}$ does annihilate the identity (hence the superscript $A$), unlike the 
$\cQ^{\epsilon}$ defined in~(\ref{eq:O}). By definition~(\ref{eq:FR}) of $\cQ^A_{\epsilon}$, 
we have 
\begin{equation}
2(\langle\cI |\cQ^A_{\epsilon})|\phi\rangle=\left\langle f_{\cI}\circ(\cQ^{\epsilon}-I\circ
\cQ^{\epsilon})\ f_{\cI}\circ\phi(0)\right\rangle=0, \label{eq:FT}
\end{equation}
because $f_{\cI}\circ I(z)=f_{\cI}(z)$ as stated in~(\ref{eq:fif}). 

\sectiono{Possible Derivation of VSFT from OSFT}\label{sec:VSFT}
We begin this section 
by reviewing the original proposal for vacuum string field theory (VSFT) made by 
Gaiotto, Rastelli, Sen and Zwiebach in refs.\cite{RSZ1,VSFT,GRSZ}. The interacting field theory 
action describing the dynamics of bosonic open strings on a single D25-brane is given by~\cite{Witten}
\begin{equation}
S_W(\Phi)=-\frac{1}{g_o^2}\left[\frac{1}{2}\langle\Phi,Q_B\Phi\rangle +\frac{1}{3}\langle\Phi,\Phi *
\Phi\rangle\right], \label{eq:A}
\end{equation}
where $g_o$ is the open string coupling constant, $Q_B$ is the BRST charge associated with the 
background of the D25-brane, $|\Phi\rangle$ is the string field represented by a state of 
ghost number 1 in the matter-ghost conformal field theory (CFT), and $\langle A,B\rangle$ denotes the 
BPZ inner product of two states $|A\rangle$ and $|B\rangle$. The precise definition of the $*$-product 
among string fields has been given in section~\ref{sec:3}. There is now strong 
evidence~\cite{LT,Rev} that the space of string field is big enough to contain the non-perturbative 
tachyon vacuum configuration which is expected to represent the closed string vacuum left after 
the unstable D-brane has disappeared. We denote by $|\Phi_0\rangle$ the tachyon vacuum configuration 
which is a solution to the equation of motion 
\begin{equation}
Q_B|\Phi_0\rangle+|\Phi_0*\Phi_0\rangle=0, \label{eq:B}
\end{equation}
although the exact form of it is still unknown. Expanding the string field about the tachyon vacuum 
as $\Phi=\Phi_0+\widetilde{\Phi}$, we can rewrite the action~(\ref{eq:A}) in the form 
\begin{equation}
\widetilde{S}_V(\widetilde{\Phi})\equiv S_W(\Phi_0+\widetilde{\Phi})-S_W(\Phi_0)=-\frac{1}{g_o^2}
\left[\frac{1}{2}\langle\widetilde{\Phi},Q\widetilde{\Phi}\rangle +\frac{1}{3}\langle\widetilde{\Phi},
\widetilde{\Phi}*\widetilde{\Phi}\rangle\right], \label{eq:C}
\end{equation}
where the new kinetic operator $Q$ has been defined by 
\begin{equation}
Q|\widetilde{\Phi}\rangle =Q_B|\widetilde{\Phi}\rangle+|\Phi_0*\widetilde{\Phi}\rangle+|\widetilde{\Phi}
*\Phi_0\rangle. \label{eq:D}
\end{equation}
Though $Q$ is expected to have vanishing cohomology, which means that there are no physical perturbative 
excitations of open strings around the tachyon vacuum, we are compelled to fall back on a numerical 
analysis in the level truncation scheme in order to investigate it directly~\cite{Noopen,FHM}. 
Instead of doing so, Rastelli, Sen and Zwiebach (henceforth RSZ) proposed in~\cite{RSZ1} that one 
should perform a field redefinition of the type 
\begin{equation}
\Psi=e^{-K}\widetilde{\Phi} \label{eq:E}
\end{equation}
which leaves the form of the cubic term unchanged, while it does alter the kinetic operator $Q$ 
into a simpler one $\cQ\equiv e^{-K}Qe^K$. RSZ conjectured that we can take $\cQ$ to be constructed 
purely out of ghost fields, and to satisfy some requisite properties. This way, we have been led 
to the following vacuum string field theory action 
\begin{equation}
S_V(\Psi)=-\frac{1}{g_o^2}\left[\frac{1}{2}\langle\Psi,\cQ\Psi\rangle +\frac{1}{3}\langle\Psi, 
\Psi *\Psi\rangle\right]. \label{eq:F}
\end{equation}
Afterward, the formulation of VSFT was completed by specifying the precise form of $\cQ$~\cite{GRSZ}. 
The authors of~\cite{GRSZ} chose a local insertion of a $c$-ghost at the string midpoint, 
\begin{equation}
\cQ=\frac{(g_o^2\kappa_0)^{1/3}}{2i}\Bigl(c(i)-c(-i)\Bigr), \label{eq:G}
\end{equation}
as a plausible candidate for the kinetic operator characterizing VSFT. It was argued that the normalization 
constant $(g_o^2\kappa_0)^{1/3}$ 
should be infinite for classical solutions to have non-vanishing action. Evidence\footnote{It 
was also shown in~\cite{GRSZ} that their choice~(\ref{eq:G}) seemed to agree numerically with another 
candidate for $\cQ$ which had been found by Hata and Kawano in~\cite{HK} by requiring that the solution 
to the equation of motion \textit{in} the Siegel gauge should also solve the full set of equations of 
motion (\textit{i.e.} without gauge-fixing).} for this special choice~(\ref{eq:G}) of $\cQ$ was given 
by explaining how such an operator could arise as a result of singular field redefinition. 
\medskip

\begin{figure}[htbp]
	\begin{center}
	\scalebox{0.92}[1]{\includegraphics{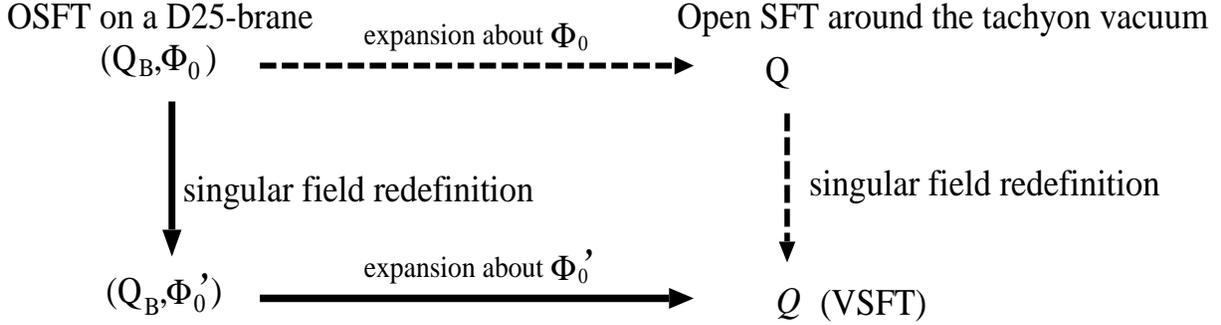}}
	\end{center}
	\caption{Each corner corresponds to the open string field theory specified by the 
	displayed kinetic operator ($Q_B,Q$ or $\cQ$). $\Phi_0$ and $\Phi_0^{\prime}$ denote 
	the tachyon vacuum configurations in the respective theories.}
	\label{fig:com}
\end{figure}

The VSFT scenario is summarized in Fig.~\ref{fig:com}. 
As described in the last paragraph, the proposal of RSZ proceeds along the dashed arrows: 
perform a field redefinition \textit{after} expanding around the tachyon vacuum solution. 
Here we assume that we can find a suitable field redefinition such that the order of these two operations 
is reversed (in other words, that this diagram is commutative). 
Moreover, suppose that the required field redefinition takes the form 
\begin{equation}
\Phi^{\prime}=e^{-K}\Phi, \label{eq:H}
\end{equation}
where the Grassmann-even operator $K$ of ghost number 0 satisfies 
\begin{eqnarray}
{}_{12}\langle R|\left(K^{(1)}+K^{(2)}\right)&=&0, \qquad (\mbox{BPZ-odd}) \label{eq:I} \\
{}_{123}\langle V_3|\left(K^{(1)}+K^{(2)}+K^{(3)}\right)&=&0, \label{eq:J} 
\end{eqnarray}
\begin{equation}
[Q_B,K]=0, \label{eq:K}
\end{equation}
in the operator language. Properties (\ref{eq:I}) and (\ref{eq:J}) guarantee that $K$ acts as a derivation 
of the $*$-algebra. Examples of such an operator are provided by 
\begin{equation}
K_n=L_n-(-1)^nL_{-n}, \label{eq:L}
\end{equation}
which generate conformal transformations that leave the string midpoint fixed. Indeed, a candidate field 
redefinition considered in~\cite{GRSZ} was induced by a reparametrization $f : \sigma \to \sigma^{\prime}$ 
of the open string coordinate $\sigma$ that satisfied $f(\pi-\sigma)=\pi-f(\sigma)$ for $0\le
\sigma\le\pi$, which implied $f(\pi/2)=\pi/2$ (midpoint fixed). 
We do not know whether every operator $K$ satisfying the requirements~(\ref{eq:I})--(\ref{eq:K}) 
can be written as a linear combination of $K_n$'s. In any case, it will suffice to postulate 
such an operator $K$ which generates the appropriate field redefinition~(\ref{eq:H}).
Thanks to the 
properties~(\ref{eq:I})--(\ref{eq:K}), the form of the original open string field theory 
action~(\ref{eq:A}) is not affected by the field redefinition~(\ref{eq:H}) at all.\footnote{Since 
it leaves the action invariant, the field redefinition~(\ref{eq:H}) may be expressed as a gauge 
transformation with a suitable choice of gauge parameter. In fact, it was shown in~\cite{FHM} and by Hata 
that $K_n$'s given in~(\ref{eq:L}) generated gauge transformations when they acted on the tachyon 
vacuum solution $\Phi_0$. Hence, in case $K$ is expressed as a linear combination of $K_n$'s, 
it follows that the new tachyon vacuum configuration $\Phi_0^{\prime}$ is actually gauge-equivalent 
to the original solution $\Phi_0$.
For our purpose, however, it does not matter whether the field 
redefinition~(\ref{eq:H}) in question belongs to gauge degrees of freedom or not.} On the other hand, 
the expression of the tachyon vacuum configuration in the new frame, denoted by $\Phi_0^{\prime}$, 
in general takes a completely different form than that $\Phi_0$ in the original frame. In particular, 
recalling that we needed a singular field redefinition to achieve the specific form~(\ref{eq:G}) 
of $\cQ$ in~\cite{GRSZ}, it is natural to think that our field redefinition~(\ref{eq:H}) should also 
be singular. Then, starting from the tachyon vacuum solution $\Phi_0$ in the original frame, 
which is expected to be regular if we remember the results obtained in the level truncation 
analysis~\cite{LT}, the new configuration $\Phi_0^{\prime}=e^{-K}\Phi_0$ must be quite singular. 
Another important property that the solution $\Phi_0^{\prime}$ must have is that the purely ghostly 
kinetic operator $\cQ$ has to arise directly from the expansion of the action around $\Phi_0^{\prime}$. 
This is because we no longer have 
degrees of freedom of field redefinition. To meet these two requirements, 
we anticipate that the solution will be of the form 
\begin{equation}
|\Phi_0^{\prime}\rangle=-Q_L|\cI\rangle+\lim_{\epsilon\to 0}\frac{a}{2}\cQ^{\epsilon}|\cI
\rangle, \label{eq:N}
\end{equation}
where $Q_L$ defined in~(\ref{eq:FC}) 
is the BRST current $j_B(z)$ integrated over left-half of the open string, 
$a$ is a normalization constant, and $\cQ^{\epsilon}$ is defined 
by\footnote{Note that our $\cQ^{\epsilon}$ is different from $\cQ_{\epsilon}$ defined 
in eq.(2.24) of \cite{GRSZ}.} 
\begin{equation}
\cQ^{\epsilon}=\frac{1}{2i}\left(e^{-i\epsilon}c(ie^{i\epsilon})-e^{i\epsilon}c(-ie^{-i\epsilon})
\right), \label{eq:O}
\end{equation}
\textit{i.e.} a $c$-ghost inserted \textit{near} the midpoint of the open string, and it smoothly 
connects to the conjectured kinetic operator $\cQ$~(\ref{eq:G}) in the limit $\epsilon\to 0$
when it is acting on a Fock space state. $\cQ^{\epsilon}$ defined 
this way is intended \textit{not} to annihilate the state $|\cI\rangle$ for any value of $\epsilon$ 
so that $\cQ^{\epsilon}|\cI\rangle$ becomes singular in the limit $\epsilon\to 0$. 
We will show that $|\Phi_0^{\prime}\rangle$ given in~(\ref{eq:N}) really solves the equation of 
motion~(\ref{eq:B}), and that this solution also possesses the second property that it gives 
rise to the desired ghostly kinetic operator $\cQ$ in a regularized form when the action~(\ref{eq:A}) 
is expanded around it. In fact, it has long been 
discussed that the configurations of the form~(\ref{eq:N}) actually solve the equation of 
motion~(\ref{eq:B}), especially in the context of purely cubic string field theory~\cite{pcsft}. 
The argument there, however, heavily relied on the existence and the formal properties of the 
identity element $\cI$ of the $*$-algebra. In this paper we have concretely defined the `identity' 
state as the wedge state of an angle $2\pi$ (see section~\ref{sec:3} for details), and have not 
required $\cI$ to satisfy $\cI *A=A*\cI =A$ for every state $A$. Since the inner product of 
$|\cI\rangle$ with any Fock space state $|\phi\rangle$ has been defined, we consider $|\cI\rangle$ not 
as a formal object but as a real one. 
In this sense, our present work may be considered as an attempt to put the arguments using the 
identity string field $\cI$ on a firm basis. 
With this in mind, we want to argue that the solution~(\ref{eq:N}) may connect the ordinary cubic 
open string field theory on a D25-brane background to vacuum string field theory. 
\medskip

From now on, we will omit the symbol $\lim_{\epsilon\to 0}$ and write simply as 
\begin{equation}
|\Phi_0^{\prime}\rangle=-Q_L|\cI\rangle+\frac{a}{2}\cQ^{\epsilon}|\cI\rangle, \label{eq:DA}
\end{equation}
with the understanding that we take 
such a limit after the entire calculation is over. First of all, we must verify that this 
configuration really solves the equation of motion 
\begin{equation}
Q_B|\Phi_0^{\prime}\rangle+|\Phi_0^{\prime}*\Phi_0^{\prime}\rangle=0. \label{eq:DB}
\end{equation}
To see this, we take the BPZ inner product of the left hand side of~(\ref{eq:DB}) with an 
arbitrary Fock space state $|\phi\rangle$. The contribution from the second term of it is 
\begin{eqnarray}
\langle\phi,\Phi_0^{\prime}*\Phi_0^{\prime}\rangle&=&\langle\phi,Q_L\cI *Q_L\cI\rangle+\frac{a^2}{4}
\langle\phi,\cQ^{\epsilon}\cI*\cQ^{\epsilon}\cI\rangle \nonumber \\
& &{}-\frac{a}{2}\left(\langle\phi,Q_L\cI *\cQ^{\epsilon}\cI\rangle+\langle\phi,
\cQ^{\epsilon}\cI *Q_L\cI\rangle\right). \label{eq:DC}
\end{eqnarray}
The second term of~(\ref{eq:DC}) vanishes because of eq.(\ref{eq:CX}). To the first term we apply 
the partial integration formulae we have derived in subsection~\ref{subsec:BRST}. On one hand, 
eq.(\ref{eq:FI}) for $\cO=Q_L$ allows us to write 
\begin{equation}
\langle\phi,Q_L\cI*Q_L\cI\rangle=\langle\phi,Q_RQ_L\cI*\cI\rangle. \label{eq:DD}
\end{equation}
On the other hand, thanks to the relation\footnote{Although we do not give explicit proofs here, 
we can show not only $\langle\cI |Q_B|\phi\rangle=0$ (\ref{eq:FD}) but also $\langle\phi,Q_B
\cI *Q_L\cI\rangle=0, \langle\phi,Q_B\cI *\cQ^{\epsilon}\cI\rangle=0,$ and $\langle\phi,Q_B\cI
*\psi\rangle=0,$ which will be sufficient for later purposes, by considering the integration 
contour of $Q_B$ just as in subsection~\ref{subsec:BRST}.}
$\langle\cI |Q_B=\langle\cI |(Q_L+Q_R)=0$ and 
eq.(\ref{eq:FJ}) for $\cO=Q_L$ we get 
\begin{equation}
\langle\phi,Q_L\cI*Q_L\cI\rangle=-\langle\phi,Q_R\cI*Q_L\cI\rangle=\langle\phi,\cI*Q_LQ_L\cI\rangle.
\label{eq:DE}
\end{equation}
Using eq.(\ref{eq:CJ}) to treat $\cI$ as the identity under the $*$-product, (\ref{eq:DD})+(\ref{eq:DE})
becomes
\[ 2\langle\phi,Q_L\cI*Q_L\cI\rangle=\langle\phi,Q_BQ_L\cI\rangle=-\langle\phi,Q_LQ_B\cI\rangle=0 \]
because $Q_B$ anticommutes with $Q_L$ (\ref{eq:FB}) and $Q_B$ kills the identity. Therefore the first 
term of~(\ref{eq:DC}) vanishes as well. Utilizing the partial integration formulae again, the second
line of eq.(\ref{eq:DC}) can be taken to the form 
\begin{eqnarray*}
-\langle\phi,Q_R\cI*\cQ^{\epsilon}\cI\rangle+\langle\phi,\cQ^{\epsilon}\cI*Q_L\cI\rangle&=&
\langle\phi, \cI*Q_L\cQ^{\epsilon}\cI\rangle+\langle\phi,Q_R\cQ^{\epsilon}\cI*\cI\rangle \\
&=&\langle\phi,Q_B\cQ^{\epsilon}\cI\rangle, 
\end{eqnarray*}
where eqs.(\ref{eq:FI}), (\ref{eq:FJ}), (\ref{eq:CJ}) have been used. Thus we have found 
\begin{equation}
\langle\phi,\Phi_0^{\prime}*\Phi_0^{\prime}\rangle=-\frac{a}{2}\langle\phi,Q_B\cQ^{\epsilon}
\cI\rangle. \label{eq:DF}
\end{equation}
The first term of~(\ref{eq:DB}) gives 
\begin{eqnarray}
\langle\phi,Q_B\Phi_0^{\prime}\rangle&=&\left\langle\phi,Q_B\left(-Q_L\cI+\frac{a}{2}\cQ^{\epsilon}
\cI\right)\right\rangle=\langle\phi,Q_LQ_B\cI\rangle+\frac{a}{2}\langle\phi,Q_B\cQ^{\epsilon}
\cI\rangle \nonumber \\ &=&\frac{a}{2}\langle\phi,Q_B\cQ^{\epsilon}\cI\rangle. \label{eq:DG}
\end{eqnarray}
Adding (\ref{eq:DF}) and (\ref{eq:DG}), we have at last reached 
\begin{equation}
\langle\phi,Q_B\Phi_0^{\prime}+\Phi_0^{\prime}*\Phi_0^{\prime}\rangle=0, \label{eq:MOE}
\end{equation}
which means that the equation of motion~(\ref{eq:DB}) is indeed solved by the 
classical configuration~(\ref{eq:DA}), at least in a weak sense. Namely, though 
we have shown that the left hand side of eq.(\ref{eq:DB}) has vanishing inner product with 
\textit{any Fock space state} in a narrow sense, we have not examined 
whether it does vanish for \textit{all} 
states including the identity state and so on. (\textit{cf}. \cite{RSZ6}) Even if the answer 
turns out to be no, we will take the standpoint that we may consider the state which solves the 
equation of motion \textit{only} in a weak sense to be a respectable solution. 
\smallskip 

We want to call the reader's attention to the fact that if the coefficient of $Q_L\cI$ in~(\ref{eq:DA}) 
were not $-1$, the cancellation between~(\ref{eq:DF}) and (\ref{eq:DG}) would be incomplete, 
accordingly the equation of motion~(\ref{eq:DB}) would not be satisfied. Thus the equation of motion 
requires this coefficient to be $-1$, as long as $a\neq 0$. To the contrary, the value of $a$ (the 
coefficient of $\cQ^{\epsilon}\cI$ ) remains completely undetermined by the equation of motion. 
This issue will be discussed in section~\ref{sec:disc}. 

\medskip

Now that we have shown $|\Phi_0^{\prime}\rangle$ given in~(\ref{eq:DA}) is a classical solution, 
let us expand the string field $\Phi^{\prime}$ around this solution as $\Phi^{\prime}=\Phi_0^{\prime}
+\Psi$ and substitute it into the action~(\ref{eq:A}) with $\Phi$ replaced by $\Phi^{\prime}$. This gives 
\[ S_W(\Phi^{\prime}_0+\Psi)=-\frac{1}{g_o^2}\sum_{n=0}^3S_n, \]
where 
\begin{eqnarray}
S_0&=&-g_o^2S_W(\Phi_0^{\prime}) \label{eq:EA} \\
&=&\frac{a^2}{8}\langle\cQ^{\epsilon}\cI,Q_B\cQ^{\epsilon}\cI\rangle+\frac{1}{2}\langle Q_L\cI , 
Q_BQ_L\cI\rangle-\frac{a}{2}\langle\cQ^{\epsilon}\cI, Q_BQ_L\cI\rangle \nonumber \\
& &{}-\frac{1}{3}\langle Q_L\cI, 
Q_L\cI *Q_L\cI\rangle+\frac{a}{2}\langle\cQ^{\epsilon}\cI, Q_L\cI *Q_L\cI\rangle \nonumber \\
& &{}-\frac{a^2}{4}\langle Q_L\cI,\cQ^{\epsilon}\cI*\cQ^{\epsilon}\cI\rangle+\frac{a^3}{24}\langle\cQ^{\epsilon}
\cI,\cQ^{\epsilon}\cI *\cQ^{\epsilon}\cI\rangle, \nonumber \\
S_1&=&\langle\Psi, Q_B\Phi_0^{\prime}+\Phi_0^{\prime}*\Phi_0^{\prime}\rangle, \label{eq:EB} \\
S_2&=&\frac{1}{2}\langle\Psi,Q_B\Psi\rangle-\langle Q_L\cI,\Psi *\Psi\rangle+\frac{a}{2}\langle
\cQ^{\epsilon}\cI,\Psi *\Psi\rangle, \label{eq:EC} \\
S_3&=&\frac{1}{3}\langle\Psi,\Psi*\Psi\rangle. \label{eq:ED}
\end{eqnarray}
If we restrict the fluctuation 
$\Psi$ to lying in the Fock space\footnote{Due to this restriction, the situation might become 
subtle when we consider non-perturbative classical solutions in the theory expanded around 
$\Phi_0^{\prime}$.}, then the term $S_1$ which is linear in $\Psi$ vanishes because of the 
equation of motion~(\ref{eq:MOE}). The term $S_2$ quadratic in $\Psi$ can be arranged as 
\begin{eqnarray}
S_2&=&\frac{1}{2}\left(\langle\Psi,Q_B\Psi\rangle-\langle\Psi,Q_L\cI *\Psi\rangle-\langle
\Psi,\Psi *Q_L\cI\rangle\right) \label{eq:EE} \\
& &{}+\frac{a}{4}\left(\langle\Psi,\Psi *\cQ^{\epsilon}\cI\rangle+\langle\Psi, \cQ^{\epsilon}
\cI *\Psi\rangle\right). \label{eq:EF}
\end{eqnarray}
Using the relation $\langle\Psi,Q_L\cI *\Psi\rangle=-\langle\Psi,Q_R\cI *\Psi\rangle$ 
and the partial integration formulae~(\ref{eq:FK}) and 
(\ref{eq:FL}), twice the first line~(\ref{eq:EE}) becomes 
\begin{eqnarray*}
& &\langle\Psi,Q_B\Psi\rangle-\langle\Psi,\cI *Q_L\Psi\rangle-\langle\Psi,Q_R\Psi*\cI\rangle
\\ &=&\langle\Psi,Q_B\Psi\rangle-\langle\Psi,(Q_L+Q_R)\Psi\rangle=0, 
\end{eqnarray*}
where we have used the relation~(\ref{eq:CA}) which allows us to treat $\cI$ as the identity 
element under the $*$-product. On the other hand, the second line~(\ref{eq:EF}) is nothing 
other than the expression~(\ref{eq:FS}). Thus we have found 
\begin{equation}
S_2=\frac{a}{2}\langle\Psi,\cQ_{\epsilon}^A\Psi\rangle. \label{eq:EG}
\end{equation}
Collecting the above results, the new action describing the theory around $\Phi_0^{\prime}$ 
becomes 
\begin{equation}
\overline{S}_V(\Psi)\equiv S_W(\Phi_0^{\prime}+\Psi)-S_W(\Phi_0^{\prime})=-\frac{1}{g_o^2}
\left[\frac{a}{2}\langle\Psi,\cQ_{\epsilon}^A\Psi\rangle+\frac{1}{3}
\langle\Psi,\Psi *\Psi\rangle\right]. \label{eq:EH}
\end{equation}
Given that the fluctuation field $\Psi$ lives in the Fock space, $\epsilon\to 0$ limit 
can be taken smoothly. Defining $\displaystyle \lim_{\epsilon\to 0}a\cQ_{\epsilon}^A
\equiv\cQ$ with an unconventional normalization factor, the above action does seem to agree 
with the conjectured VSFT action~(\ref{eq:F}). Moreover, the kinetic operator $\cQ$ has arisen 
with being regularized in such a way that it should annihilate the identity state $\cQ |\cI\rangle
=\lim_{\epsilon\to 0} a\cQ_{\epsilon}^A|\cI\rangle=0$, as shown in~(\ref{eq:FT}). 
\smallskip

To sum up, what we have shown is that the classical configuration~(\ref{eq:N}) gives a 
solution to the equation of motion~(\ref{eq:DB}) derived from the action~(\ref{eq:A}) 
describing the original D25-brane, and that the string field theory action expanded 
around this solution coincides with the vacuum string field theory action proposed 
in~\cite{GRSZ}, \textit{up to} the normalization of the kinetic operator. 

\sectiono{Discussion}\label{sec:disc}
In this paper, we have focused our attention mainly on the conformal field 
theory description of the identity string field $\cI$. By giving the identity state 
a precise definition as the wedge state of an angle $2\pi$, we have avoided falling into 
formal arguments which have been given so far. Furthermore, the generalized gluing and 
resmoothing theorem has provided us with a useful computational tool which is suitable 
for dealing with the wedge-type surface states. Based on the results obtained from the 
CFT calculations, we have specified a solution around which the theory is described by 
the vacuum string field theory action proposed in~\cite{GRSZ}. We guess that this solution 
represents the tachyon vacuum in some singular coordinate system. In the rest of this concluding 
section we discuss some open questions and give some remarks on further studies. 

\subsection{Uniqueness of the solution?}
In the previous sections, we have shown that the classical configuration 
\begin{equation}
\Phi_0^{\prime}(t,a)=-tQ_L\cI+\frac{a}{2}\cQ^{\epsilon}\cI \label{eq:XA}
\end{equation}
solves the equation of motion~(\ref{eq:DB}) if $t=1$. As remarked in section~\ref{sec:VSFT}, 
however, the equation of motion does not constrain the value of $a$ at all. This seems 
problematic because we then have a one-parameter family of solutions, which prevents us from 
interpreting $\Phi_0^{\prime}(1,a)$ as the tachyon vacuum solution. Although we have no 
appropriate way of determining the value of $a$ uniquely, it would be worthwhile to notice 
that in order for the action~(\ref{eq:EH}) around $\Phi_0^{\prime}$ to agree with the 
VSFT action~(\ref{eq:F}) even including their overall normalizations, we must have 
$a=(g_o^2\kappa_0)^{1/3}$. Since there is an argument suggesting that one needs to take $\kappa_0$ to 
be infinite, we shall tentatively assume that only $\Phi_0^{\prime}(t=1,a=\infty)$ corresponds to 
the tachyon vacuum solution. 

\begin{figure}[htbp]
	\begin{center}
	\scalebox{0.7}[0.7]{\includegraphics{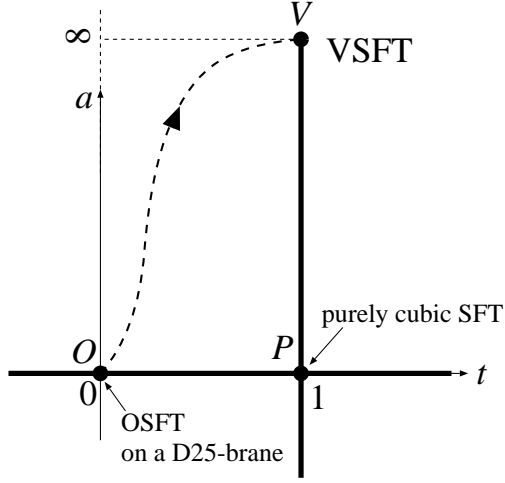}}
	\end{center}
	\caption{Two-dimensional configuration space of $\Phi_0^{\prime}$~(\ref{eq:XA}). At 
	every point on the bold lines $\Phi_0^{\prime}(t,a)$ gives a solution to the equation 
	of motion.}
	\label{fig:taspace}
\end{figure}

In addition, if we set $a=0$, $\Phi_0^{\prime}(t,0)=-tQ_L\cI$ 
gives a solution to the equation of motion~(\ref{eq:DB}) for any value of $t$. This is because 
each term of the left hand side of~(\ref{eq:DB}) vanishes independently of each other 
according to the argument given in section~\ref{sec:VSFT}. Note that $(t,a)=(1,0)$ leads to 
the purely cubic string field theory~\cite{pcsft}. The configuration space of $\Phi_0^{\prime}$ 
parametrized by $(t,a)$ is drawn in Fig.~\ref{fig:taspace}. 
We see that the set of all solutions of the form~(\ref{eq:XA}) consists of two branches 
which meet at the purely cubic string field theory point. 
\medskip

Let us consider the tachyon condensation process which is now supposed to roll down from 
$(t,a)=(0,0)$ to $(1,\infty)$. Since it sounds unlikely that the string field $\Phi^{\prime}$ 
keeps on satisfying the equation of motion during the condensation process, we assume that 
it proceeds along the dashed curve $OV$ shown in Fig.~\ref{fig:taspace}, rather than along 
the bold lines $OPV$. In the vicinity of the VSFT point, we introduce a new parameter $\Lambda$ 
through the relation 
\begin{equation}
t=1-\frac{a}{\Lambda}. \label{eq:XB}
\end{equation}
Of course, $t=1$ corresponds to $\Lambda\to\infty$. On the dashed curve $OV$, $t$ and $a$ are 
implicitly determined as functions of $\Lambda$, so we write them as $t=t(\Lambda), 
a=a(\Lambda)$. Hence we can think of $\Lambda$ as the coordinate on the curve $OV$. 
Now consider the configuration $\Phi_0^{\prime}\left(1-\frac{a(\Lambda)}{\Lambda},
a(\Lambda)\right)$ and expand the original action~(\ref{eq:A}) around it, even though it 
\textit{does not} correspond to a classical solution. The resulting action is found to be 
\begin{eqnarray}
\widehat{S}_V(\Psi;\Lambda)\equiv S_W(\Phi_0^{\prime}+\Psi)-S_W(\Phi_0^{\prime})&=&
-\frac{1}{g_o^2}\Biggl[ \frac{a(\Lambda)^2}{2\Lambda}\langle\Psi,Q_B\cQ^{\epsilon}\cI\rangle
+\frac{a(\Lambda)}{2\Lambda}\langle\Psi,Q_B\Psi\rangle \nonumber \\
& &{}+\frac{a(\Lambda)}{2}\langle\Psi,\cQ_{\epsilon}^A\Psi\rangle+\frac{1}{3}
\langle\Psi,\Psi *\Psi\rangle\Biggr]. \label{eq:XC}
\end{eqnarray}
Below, we simply ignore the first term which is linear in $\Psi$ and would not exist if 
$\Phi_0^{\prime}$ were a solution. Upon fixing in the Siegel gauge, the remaining terms 
are written as 
\begin{equation}
\widehat{S}_V(\Psi;\Lambda)=-\frac{1}{g_o^2}\left[\frac{a(\Lambda)}{2}\left\langle\Psi,c_0
\left(1+\frac{L_0^{\mathrm{tot}}}{\Lambda}\right)\Psi\right\rangle+\frac{1}{3}\langle\Psi,
\Psi *\Psi\rangle\right]. \label{eq:XD}
\end{equation}
This form closely resembles the `regularized VSFT action' considered in~\cite{GRSZ} to 
regularize the singular nature of vacuum string field theory. As the authors of~\cite{GRSZ} 
mentioned, by taking the regularization parameter $\Lambda$ to be infinite (hence $t=1$), 
$\widehat{S}_V$ reduces to the singular VSFT action~(\ref{eq:EH}) in a gauge-fixed form. 
There is, however, a wide difference between ours~(\ref{eq:XD}) and theirs. Their regularization 
is achieved by replacing the singular reparametrization with a nearly singular one leading to 
an equivalent theory, whereas our regularization represented by $\Lambda$ corresponds to 
the deformation of the classical configuration around which we are going to expand the action, 
which of course is expected to give rise to an inequivalent theory. In particular, we have to 
take into account the term linear in $\Psi$ since for finite $\Lambda$,  $\Phi_0^{\prime}$ is 
not even a solution. Anyway, it is obvious that we need a better understanding of the 
regularization procedure to find a precise relation between ours and theirs. 

\subsection{D25-brane tension}
\begin{figure}[htbp]
	\begin{center}
	\scalebox{0.7}[0.7]{\includegraphics{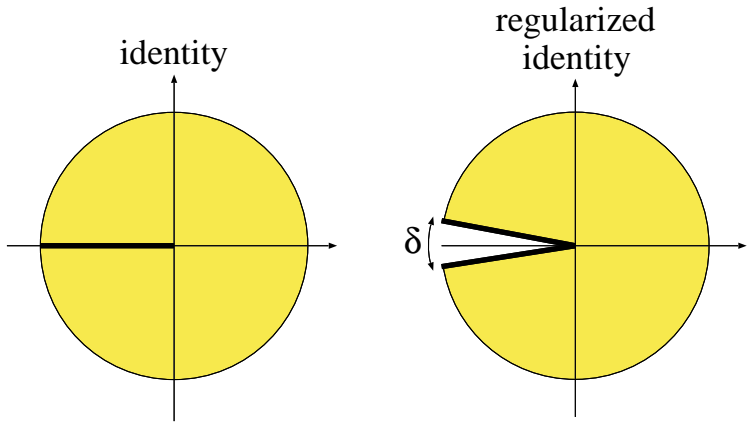}}
	\end{center}
	\caption{Deformation of the identity state.}
	\label{fig:pacman}
\end{figure}

In spite of the fact that we started from the well-established cubic open string field theory 
on a D25-brane, we are unable to calculate the energy density associated with the solution. 
This problem is closely related to the issue mentioned in the last subsection, namely 
that the parameter $a$ is left unfixed. Thinking in reverse order, it might be possible 
to determine $a$ by requiring that the value of the action $S_W(\Phi_0^{\prime}(t=1,a))$ 
be equal to the negative of the energy of a single D25-brane, $-V_{26}\cT_{25}$, where 
$V_{26}$ represents the volume of the 26-dimensional spacetime and $\cT_{25}$ denotes 
the tension of the D25-brane. Using the relation~\cite{Univ} 
\begin{equation}
\cT_{25}=\frac{1}{2\pi^2g_o^2}, \label{eq:XE}
\end{equation}
the equation we want to solve becomes 
\begin{equation}
-g_o^2S_W(\Phi_0^{\prime}(1,a))\equiv S_0(a)=\frac{V_{26}}{2\pi^2}, \label{eq:XF}
\end{equation}
where the expression for $S_0(a)$ has been given in eq.(\ref{eq:EA}). If we could solve 
this equation with respect to $a$, the result of which is expected to diverge in some sense, 
then there would remain no unfixed parameter in the theory around $\Phi_0^{\prime}$ 
(and presumably in VSFT). But unfortunately, it has turned out that $S_0(a)$ is quite 
badly-behaved. For example, we have obtained\footnote{
	Note that this quantity would vanish if we na\"{\i}vely use equation of motion for $\delta=0$.
}
\begin{equation}
\langle {\cal Q}^{\epsilon}{\widetilde {\cal I}}_{\delta}, Q_B  {\cal Q}^{\epsilon}
{\widetilde{\cal I}}_{\delta} \rangle = -\delta^2\sin^2\epsilon\left[\frac{1}{2}\left\{\left(
\tan{\epsilon\over2}\right)^{2\over\delta}+\left(\tan{\epsilon\over2}\right)^{-{2\over\delta}}\right\}+3
\right]V_{26}, \label{eq:XG}
\end{equation}
where $\delta$ is a regularization parameter introduced to deform the identity state in a way 
indicated in Fig.~\ref{fig:pacman}. 
To be more specific, the regularized identity state is defined by 
\begin{equation}
\left\langle\widetilde{\cI}_{\delta},\cO\right\rangle=\left\langle\widetilde{f}_{\cI_{\delta}}
\circ\cO(0)\right\rangle \quad\  \mathrm{with}\quad \widetilde{f}_{\cI_{\delta}}(z)=h^{-1}
\left(h(z)^{\frac{2}{1+\delta}}\right). \label{eq:XH2}
\end{equation}
This regularization is necessary because if we na\"{\i}vely apply the gluing theorem to 
the expression of the form $\langle\cI,\cI\rangle$ then we are left with an ill-defined 
expression. This can be seen from the fact that we cannot take the $\delta\to 0$ limit 
in eq.(\ref{eq:XG}). The evaluation of $S_0(a)$ has not been successful up to now, 
and is under investigation~\cite{KMO}. 

\subsection{Other solutions?}
Since the action $\overline{S}_V(\Psi)$ in~(\ref{eq:EH}) has been obtained by expanding 
the original action $S_W(\Phi^{\prime})$~(\ref{eq:A}) around the classical solution 
$\Phi_0^{\prime}=-Q_L\cI+\frac{a}{2}\cQ^{\epsilon}\cI$, we can recover the action 
$S_W(\Phi^{\prime})$ by re-expanding the `VSFT action' $\overline{S}_V(\Psi)$ around 
the solution 
\begin{equation}
\Psi_0=-\Phi_0^{\prime}=Q_L\cI-\frac{a}{2}\cQ^{\epsilon}\cI \label{eq:XH}
\end{equation}
as $\Psi=\Psi_0+\Phi^{\prime}$, and it is obvious that this solution represents a single 
D25-brane. As remarked in a footnote of section~\ref{sec:VSFT}, however, the action 
$\overline{S}_V(\Psi)$ given in~(\ref{eq:EH}) has been shown to be valid only for $\Psi$ 
lying in the Fock space. Since the above solution $\Psi_0$ is outside the Fock space in a 
narrow sense, it may not be correct to suppose the theory around $\Phi_0^{\prime}$ to be 
literally described by the action $\overline{S}_V(\Psi)$ in this case. Nevertheless, 
we shall have arguments below on the assumption that the theory around $\Phi_0^{\prime}$ 
is really governed by $\overline{S}_V(\Psi)$. 
\medskip 

If we are to regard $\overline{S}_V(\Psi)$ as representing the vacuum string field theory 
under an appropriate choice of the normalization constant $a$, then we must be able to 
construct other D-brane solutions than the original single D25-brane, that is to say, 
lower-dimensional D-branes, multiple D-branes, and so forth. However, as pointed out 
in~\cite{RSZ2}, it seems difficult to produce spatially localized solutions if we adhere 
to the identity-type solution of the form~(\ref{eq:XH}). It is true that once a D25-brane~(\ref{eq:XH}) 
has been constructed lower-dimensional D-branes could be obtained as lump solutions on 
the D25-brane via tachyon condensation at least in the level truncation 
scheme~\cite{lump}, but it is obviously desirable to build them directly. 
\smallskip

In this connection, we also give a brief comment on the possible relation to the 
projector-type\footnote{By the word `projector-type solutions' we mean the (twisted) sliver state and 
the (twisted) butterfly state as well as the family they belong to.} solutions which have been 
examined in the context of vacuum string field theory~\cite{RSZ2,GRSZ}. 
If both the identity-type solution~(\ref{eq:XH}) and one of the spatially-independent projector-type 
solutions represent one and only D25-brane, they should be related to each other through some 
gauge transformation. We do not know whether such a surprising thing could happen. 

\subsection{Mystery on $c_0\cI$ revisited}
The following is a well-known argument: Since $c_0$ acts as a derivation of the $*$-algebra, 
it follows that 
\begin{equation}
c_0A=c_0(\cI *A)=(c_0\cI)*A+\cI *(c_0A)=(c_0\cI)*A+c_0A \label{eq:XI}
\end{equation}
for a \textit{true identity state} $\cI$ and for any $A$. For this equation to hold, we must 
have $c_0\cI=0$, which is not the case~\cite{RZ}. Hence one is tempted to conclude that there 
does not exist such a true identity state. Though this conclusion may be true in a sense, the 
above argument is too strong. In fact, it is sufficient to have $(c_0\cI)*A=0$ rather than 
$c_0\cI=0$. That is to say, there is no problem in eq.(\ref{eq:XI}) if we show that $c_0\cI$ 
gives zero whenever it is $*$-multiplied by an arbitrary element. In the operator language, 
this statement is equivalent to the relation ${}_1\langle\cI |c_0^{(1)}|V_3\rangle_{123}=0$, and 
it was shown in~\cite{Kishimoto} that it is indeed true. 
\smallskip

From the point of view of our present paper, we can calculate 
\begin{equation}
\langle\phi,(c_0\cI)*A\rangle+\langle\phi,\cI *(c_0A)\rangle \label{eq:XJ}
\end{equation}
and see if it is equal to $\langle\phi,c_0A\rangle$. For simplicity we take $|\phi\rangle$ and 
$|A\rangle$ to be Fock space states. The second term of~(\ref{eq:XJ}) reduces to $\langle 
\phi,c_0A\rangle$ thanks to the relation~(\ref{eq:CA}). We have calculated the first term 
with the help of the gluing theorem and obtained 
\begin{equation}
\langle\phi,(c_0\cI)*A\rangle =\langle\phi,c_0A\rangle\neq 0, \label{eq:XK}
\end{equation}
contrary to our expectation. In addition, we can easily show $\langle\phi,c_0(\cI *A)\rangle
=\langle\phi,c_0A\rangle$ directly. After all, we have found 
\begin{equation}
\langle\phi,(c_0\cI)*A\rangle+\langle\phi,\cI *(c_0A)\rangle=2\langle\phi,c_0A\rangle
\neq \langle\phi,c_0A\rangle=\langle\phi,c_0(\cI *A)\rangle. \label{eq:XL}
\end{equation}
Since there is no ambiguity in our computational scheme, something must be wrong with 
eq.(\ref{eq:XI}). A possible resolution of this problem would be to think that it depends on 
the states to be acted on whether the operator $c_0$ may be regarded as a derivation. 
We have not reached a definite conclusion concerning this point. 
To have a better understanding of the identity state, it would be important to examine 
the action of such outer derivations on $\cI$ in more detail. 

\section*{Acknowledgements}
We are grateful to Yutaka Matsuo for his collaboration at various stages of this work 
and for suggesting the usefulness of identity string field to us. 
I.~K. would like to thank H.~Hata, T.~Kugo and S.~Moriyama for helpful
discussions.
K. O. wants to express his gratitude to T. Eguchi for discussion and 
to C. Devens for encouragements. 





\end{document}